\documentclass[11pt,a4paper,longbibliography]{article}
\usepackage{jcappub} 

\usepackage{graphicx} 
\usepackage[utf8]{inputenc}
\usepackage{subcaption}
\usepackage{physics}
\usepackage{amsfonts,amssymb,amsmath}
\usepackage{mathtools} 
\usepackage[T1]{fontenc}
\usepackage{tikz}
\usetikzlibrary{positioning, shapes.geometric, arrows.meta, fit, calc, decorations.markings}
\usepackage{diagbox}
\usepackage{mathtools} 
\usepackage{multirow}
\usepackage{makecell} 

\usepackage{xcolor}
\newcommand{\change}[1]{#1}

\newcommand{\hyperparams}{\Lambda}
\newcommand{\hyperpowerspectrum}{\Lambda_{\rm PS}}
\newcommand{\hyperGalaxies}{\Lambda_{\rm g}}
\newcommand{\hypermagnitudes}{\Lambda_{\rm M}}

\newcommand{\hypercosmology}{\Lambda_{\rm c}}

\newcommand{\rhogbar}{\overline{ \rho_{\rm g}}}
\newcommand{\deltagexp}{\delta_{\rm g,exp}}
\newcommand{\epsilong}{\epsilon_{\rm g}}

\newcommand{\deffrom}{\coloneqq}
\newcommand{\p}{\mathbb{P}}
\newcommand{\deltaDirac}{\delta_{\rm D}}

\newcommand{\Omegam}{\Omega_{{\rm m}}}
\newcommand{\omegacdm}{\omega_{{\rm cdm}}}
\newcommand{\omegab}{\omega_{{\rm b}}}
\newcommand{\hu}{\,{\rm km \,s^{-1} \, Mpc^{-1}}} %
\newcommand{\Mpc}{\,{\rm Mpc}} %

\newcommand{\rhoDM}{\rho_{{\rm m}}}
\newcommand{\rhoDMbar}{\overline{\rho_{{\rm m}}}}
\newcommand{\densityDM}{\delta_{\rm DM}}
\newcommand{\densityDMpixel}{\delta_{{\rm DM},I}}
\newcommand{\densityDMpixelall}{\delta_{{\rm DM},I\in\mathcal{I}}}
\newcommand{\samples}{\mathcal{S}}
\newcommand{\powerspectrumDM}{P_{\rm DM}^{(K)}}
\newcommand{\densityDMFourier}{\delta_{\rm DM}^{(K)}}

\newcommand{\countGalaxiesObs}{n_{\rm c}}

\newcommand{\countGalaxiesNonObspixel}{n_{\overline{{\rm c}},I}}

\newcommand{\countGalaxiesObspixel}{n_{{\rm c},I}}

\newcommand{\rateGalaxies}{\mu(z,\skypos|\hyperparams, \densityDM)}

\newcommand{\countGalaxiesTrue}{n_{\rm g}}
\newcommand{\countGalaxiesTruepixel}{n_{{\rm g},I}}
\newcommand{\countGalaxiesTruepixelall}{n_{{\rm g},I\in\mathcal{I}}}
\newcommand{\countGalaxiesTrueEstimator}{\hat{n}_{\rm g}}
\newcommand{\countGalaxiesTrueEstimatorpixel}{\hat{n}_{{\rm g},I}}
\newcommand{\mthresh}{m_{\rm thr}}
\newcommand{\Mthresh}{M_{\rm thr}}

\newcommand{\fieldgausswhitenedspatial}{G_{\rm w}}

\newcommand{\fieldgausswhitenedfourier}{G^{(K)}_{\rm w}}
\newcommand{\fieldgausscoloredfourier}{G^{(K)}_{\rm c}}
\newcommand{\fieldgausscoloredspatial}{G_{\rm c}}

\newcommand{\Poisson}{\mathrm{Poisson}}
\newcommand{\Binomial}{\mathrm{Binomial}}

\newcommand{\Deltarelpixel}{\Delta_{{\rm rel},I}}
\newcommand{\Deltapixel}{\Delta_{I}}

\newcommand{\PSamplitude}{A_{\rm phen, S}}

\newcommand{\PSalpha}{\alpha_{\rm phen, S}}
\newcommand{\PSn}{n_{\rm phen, S}}

\newcommand{\spatialcorrelationfieldgaussian}{\xi_{G_c}}
\newcommand{\spatialcorrelationfieldlognormal}{\xi_{\rm LN}}

\newcommand{\pdetpixel}{p_{{\rm det}, I}}

\newcommand{\pnotdetpixel}{p_{{\rm not\,det}, I}}

\newcommand{\numpyro}{\texttt{numpyro}}
\newcommand{\cosmopowerjax}{\texttt{CosmoPower-jax}}

\newcommand{\cmin}{c_{\rm min }}
\newcommand{\etazero}{\eta_{0}}

\newcommand{\muThresholdAbove}{M_{f}}
\newcommand{\muFraction}{f_{M}}

\newcommand{\skypos}{\varphi_{\perp}}

\newcommand{\Data}{\mathcal{D}}
\newcommand{\Datalist}[1]{\{\Data\}_{#1}}

\newcommand{\rateAll}[1]{\mu(\Data_{#1}|\hyperparams, \densityDM)}

\newcommand{\rateAbspixel}{\mu_{{\rm a},I}}

\newcommand{\rateAllpixel}[1]{\mu_I(\Data_{#1}|\hyperparams, \densityDM)}

\newcommand{\rateAlldet}[1]{\mu_{{\rm det}}(\Data_{#1}|\hyperparams, \densityDM)}
\newcommand{\rateAlldetpixel}[1]{\mu_{{\rm det},I}(\Data_{#1}|\hyperparams, \densityDM)}
\newcommand{\rateAllnondetpixel}[1]{\mu_{{\rm not\;det},I}(\Data_{#1}|\hyperparams, \densityDM)}

\newcommand{\finiteData}[1]{\Delta\Data_{#1}}

\newcommand{\overcount}{{\rm OC}[\{n_i\}_{i=1,2,\ldots,N}]}
\usepackage[acronym, toc, nonumberlist]{glossaries}

\glsdisablehyper
\setacronymstyle{long-short}

\newacronym{ns}{NS}{neutron star}
\newacronym{bh}{BH}{black hole}
\newacronym{bbh}{BBH}{binary black hole}
\newacronym{bns}{BNS}{binary neutron star}
\newacronym{nsbh}{NSBH}{neutron star black hole}

\newacronym{eos}{EoS}{equation of state}
\newacronym{gw}{GW}{gravitational wave}
\newacronym{gr}{GR}{general relativity}
\newacronym{snr}{SNR}{signal-to-noise ratio}

\newacronym{lisa}{LISA}{Laser Interferometer Space Antenna }
\newacronym{ligo}{LIGO}{Laser Interferometer Gravitational wave Observatory}
\newacronym{kagra}{KAGRA}{KAmioka GRavitational wave detector}
\newacronym{eob}{EOB}{effective one-body}
\newacronym{em}{EM}{electromagnetic}
\newacronym{lcdm}{$\Lambda$CDM}{$\Lambda$ cold dark matter}
\newacronym{pl}{PL}{power law}
\newacronym{plg}{PLG}{power law and Gaussian}
\newacronym{kde}{KDE}{kernel density estimate}
\newacronym{de}{DE}{dark energy}
\newacronym{cdf}{CDF}{cumulative density function}

\newacronym{lvk}{LVK}{LIGO-Virgo-KAGRA}
\newacronym{ego}{EGO}{European gravitational observatory}
\newacronym{asd}{ASD}{amplitude spectral density}
\newacronym{psd}{PSD}{power spectral density}
\newacronym{mcmc}{MCMC}{Monte Carlo Markov chain}
\newacronym{hlv}{HLV}{Hanford Livingston Virgo}
\newacronym{pe}{PE}{parameter estimation}
\newacronym{cbc}{CBC}{compact binary coalescence}
\newacronym{aligo}{aLIGO}{advanced LIGO}
\newacronym{far}{FAR}{false alarm rate}

\newacronym{cl}{CL}{confidence level}

\newacronym{pn}{PN}{post-Newtonian}
\newacronym{nr}{NR}{numerical relativity}
\newacronym{ppisn}{PPISN}{pulsation pair-instability supernova}
\newacronym{pisn}{PISN}{pair instability-supernova}
\newacronym{et}{ET}{Einstein Telescope}
\newacronym{ce}{CE}{Cosmic Explorer}

\newacronym{cmb}{CMB}{cosmic microwave background}
\newacronym{lss}{LSS}{large scale structure}
\newacronym{isco}{ISCO}{innermost stable orbit}

\newacronym{oi}{Oi}{observation run $i$}
\newacronym{gwtci}{GWTC-i}{gravitational wave transient catalog $i$}

\newacronym{2g}{2G}{second generation}
\newacronym{3g}{3G}{third generation}

\newacronym{bao}{BAO}{baryonic acoustic oscillation}

\newacronym{wkb}{WKB}{Wentzel–Kramers–Brillouin}

\newacronym{dpg}{DPG}{Dvali-Gabadadze-Porrati}
\newacronym{dhost}{DHOST}{degenerate higher-order scalar-tensor}
\newacronym{mg}{MG}{modified gravity}
\newacronym{des}{DES}{dark energy survey}

\newacronym{tt}{TT}{transverse-traceless}

\newacronym{sgwb}{SGWB}{stochastic gravitational wave background}
\newacronym{dgrb}{DGRB}{diffuse $\gamma$-ray background}
\newacronym{gut}{GUT}{grand unified theory}

\newacronym{ng}{NG}{Nambu-Goto}
\newacronym{gbr}{GBR}{gravitational backreaction}

\newacronym{nf}{NF}{normalizing flow}
\newacronym{ml}{ML}{machine learning}
\newacronym{lfi}{LFI}{likelihood-free inference}
\newacronym{nn}{NN}{neural network}
\newacronym{dingo}{DINGO}{deep inference for gravitational wave observations}
\newacronym{gpu}{GPU}{graphics processing unit}
\newacronym{hba}{HBA}{hierarchical Bayesian analysis}

\newacronym{kl}{KL}{Kullback-Leibler}
\newacronym{js}{JS}{Jensen-Shannon}
\newacronym{ks}{KS}{Kolmogorov–Smirnov}

\newacronym{smbh}{SMBH}{supermassive black hole}
\newacronym{agn}{AGN}{active galactic nuclei}
\newacronym{spa}{SPA}{stationary phase approximation}

\newacronym{pta}{PTA}{pulsar timing array}

\newacronym{npe}{NPE}{neural posterior estimation}

\newacronym{dm}{DM}{dark matter}

\newacronym{grf}{GRF}{Gaussian random field}

\title{Cosmic Cartography: Bayesian reconstruction of the galaxy density informed by large-scale structure}
\author{}
\date{September 2024}

\author{Konstantin Leyde$^a$,}
\author{Tessa Baker$^a$,}
\author{Wolfgang Enzi$^a$}
\affiliation[a]{Institute of Cosmology and Gravitation, University of Portsmouth, \\
Burnaby Road, Portsmouth PO1 3FX, United Kingdom}
\emailAdd{konstantin.leyde@port.ac.uk}
\emailAdd{tessa.baker@port.ac.uk}
\emailAdd{wolfgang.enzi@port.ac.uk}

\begin{abstract}{
The dark sirens method combines gravitational waves and catalogs of galaxies to constrain the cosmological expansion history, merger rates and mass distributions of compact objects, and the laws of gravity. However, the incompleteness of galaxy catalogs means faint potential host galaxies are unobserved, and must be modeled to avoid inducing a bias. The majority of dark sirens analyses to date assume that the missing galaxies are distributed uniformly across the sky, which is clearly unphysical. We introduce a new Bayesian approach to the reconstruction of galaxy catalogs, which makes full use of our knowledge of large-scale structure. Our method quantifies the uncertainties on the estimated true galaxy number count in each voxel, and is marginalized over cosmological parameters and bias parameters. Crucially, our method further assesses the (absolute) magnitude distribution of galaxies, which is not known from the galaxy catalog itself. We present the details of our method and validate our approach on a galaxy catalog associated to the Millennium Simulation. The tools developed here generate physically-informed and robust host galaxy reconstructions, enabling more informative dark sirens analyses. 
Stage IV galaxy surveys will display greater redshift overlap with GW observations, whilst remaining incomplete -- emphasizing the importance of our work.
}
\end{abstract}

\begin{document}

\maketitle


\section{Introduction}
\Glspl{gw} offer novel means to measure cosmological parameters through various methods that combine the \gls{gw} data with other types of data, such as direct electromagnetic (EM) counterparts \cite{LIGOScientific:2017vwq, LIGOScientific:2017adf} or galaxy catalogs \cite{Schutz:1986gp, DelPozzo:2011vcw, Gray:2019ksv, Gray:2021sew, Finke:2021aom, Turski:2023lxq, Mastrogiovanni:2023emh, Gray:2023wgj, Perna:2024lod}, or rely on \gls{gw} data alone using the redshifted source frame mass distribution \cite{Taylor:2011fs, Taylor:2012db, Farr:2019twy, Mastrogiovanni:2021wsd, Mancarella:2021ecn, Leyde:2022orh, Ezquiaga:2022zkx, Pierra:2023deu, Leyde:2023iof, Farah:2024xub, MaganaHernandez:2024uty}. 
While \glspl{gw} with direct EM counterparts provide the most precise \gls{gw} constraints on today's expansion rate of the Universe, $H_0$, current observations indicate that such multi-messenger events are infrequent, prompting the development of alternative approaches. 
The galaxy catalog method (a.k.a. dark siren method) represents a powerful tool for an $H_0$ measurement, which can, in principle, utilize \textit{all} detected \gls{gw} signals: \gls{gw} data alone informs on the sky position, which one can combine with the galaxy catalog data. 
An assumption on the host probability for each galaxy provides redshift information and hence, when combined with the luminosity distance information from the \gls{gw}, constrains the cosmological parameters appearing in the luminosity distance-redshift relation.

Despite its potential, the galaxy catalog method is hindered by the incompleteness of available galaxy catalogs -- at higher redshifts, an increasingly larger fraction of galaxies will start to fall above the apparent magnitude limit of the survey \cite{Dalya:2018cnd, Dalya:2021ewn}.
Additionally, with the improvement of upcoming detector networks, GWs will be observed at increasingly high redshifts; unless highly-complete full-sky surveys are available, galaxy catalogs will provide on average sparser redshift information, resulting in a less informative $H_0$ measurement.  

Almost all current approaches of the galaxy catalog method address this challenge by assuming a uniform distribution in comoving volume for the \gls{gw} host if the catalog is incomplete (for exceptions see \cite{Finke:2021aom} and \cite{Dalang:2023ehp}). 
However, countless observations show that the distribution of galaxies (and thereby of \gls{gw} sources) is not uniform. 
Indeed, the distribution of galaxies shows structures such as voids, clusters, filaments, and sheets that together form the so-called cosmic web.

We propose to leverage this large-scale correlation in reconstructing the out-of-catalog (unobserved) galaxies, yielding a more informative redshift prior for the potential \gls{gw} hosts and hence, allowing one to measure $H_0$ more precisely. 
This work prepares for an improved dark siren analysis using Stage IV large-scale structure data and current GW data \cite{det1-aligo2015,det2-aLIGO:2020wna,det3-Tse:2019wcy,det4-VIRGO:2014yos,det5-Virgo:2019juy}, as opposed to future forecasts with third-generation detectors \cite{Sathyaprakash:2009xt, Chan:2018csa, Maggiore:2019uih, Evans:2021gyd, Branchesi:2023mws, Muttoni:2023prw, Chen:2024gdn}.
Our Bayesian framework enables a robust reconstruction of galaxy catalogs, capable of determining the true galaxy number count within each voxel (three-dimensional pixel) and quantifying its associated number count uncertainty. 
We incorporate the three-dimensional correlation of the \gls{dm} field, estimating the \gls{dm} power spectrum in conjunction with the missing galaxy counts. In addition to the number counts in each voxel, we use the galaxies' redshift, and apparent magnitude information.
From the latter, we estimate the absolute magnitude distribution of galaxies, which remains poorly known, and that will prove essential to estimate the missing galaxy number count. 
This approach also allows for the simultaneous variation of cosmological parameters, $H_0$ and the current matter content $\Omegam$, alongside galaxy number count, cosmological parameters, and bias parameters that link galaxy density to \gls{dm} density. 
We validate our methodology on galaxy catalogs that use the dark matter density of the Millennium Simulation, demonstrating its efficacy and potential to apply it to the \gls{gw} catalogs of the currently on-going observation run that will be available in 2025 (O4a) and 2026/2027 (O4b) \cite{ligo_observation_run_o4_2023}.

A number of works have explored forward modeling to reconstruct the initial \gls{dm} density field from an observed galaxy catalog: in particular, the Bayesian Origin Reconstruction from Galaxies (BORG) inference algorithm developed by \cite{Jasche:2009hx, Jasche:2009hz, Jasche:2009ia, Jasche:2012kq, Jasche:2013lwa} evolves initially Gaussian density fluctuations at high redshift over time, comparing the simulated light-cone data to observations. 
Typically, this framework fixes the cosmological parameters, precisely the parameters we intend to measure with \gls{gw} sources. This is with the exception of the work in \cite{Ramanah:2018eed}.
Other works also rely on forward-modeling but do not formulate an explicit likelihood, see for instance \cite{He:2018ggn, Kaushal:2021hqv, Jamieson:2022lqc, SimBIG:2023ywd, deSanti:2023rsw, Saadeh:2024vuj} that explore field-level inference with machine learning.
In the context of \gls{gw} cosmology, the work of \cite{Dalang:2023ehp, Dalang:2024gfk} has formulated a phenomenological maximum likelihood estimator for the missing galaxy number counts. This approach leverages the number count statistic into an improved reconstruction of the catalog, yielding better results than the standard completion that is uniform in comoving volume. 

\begin{figure}
    \centering
    \includegraphics[width=\textwidth]{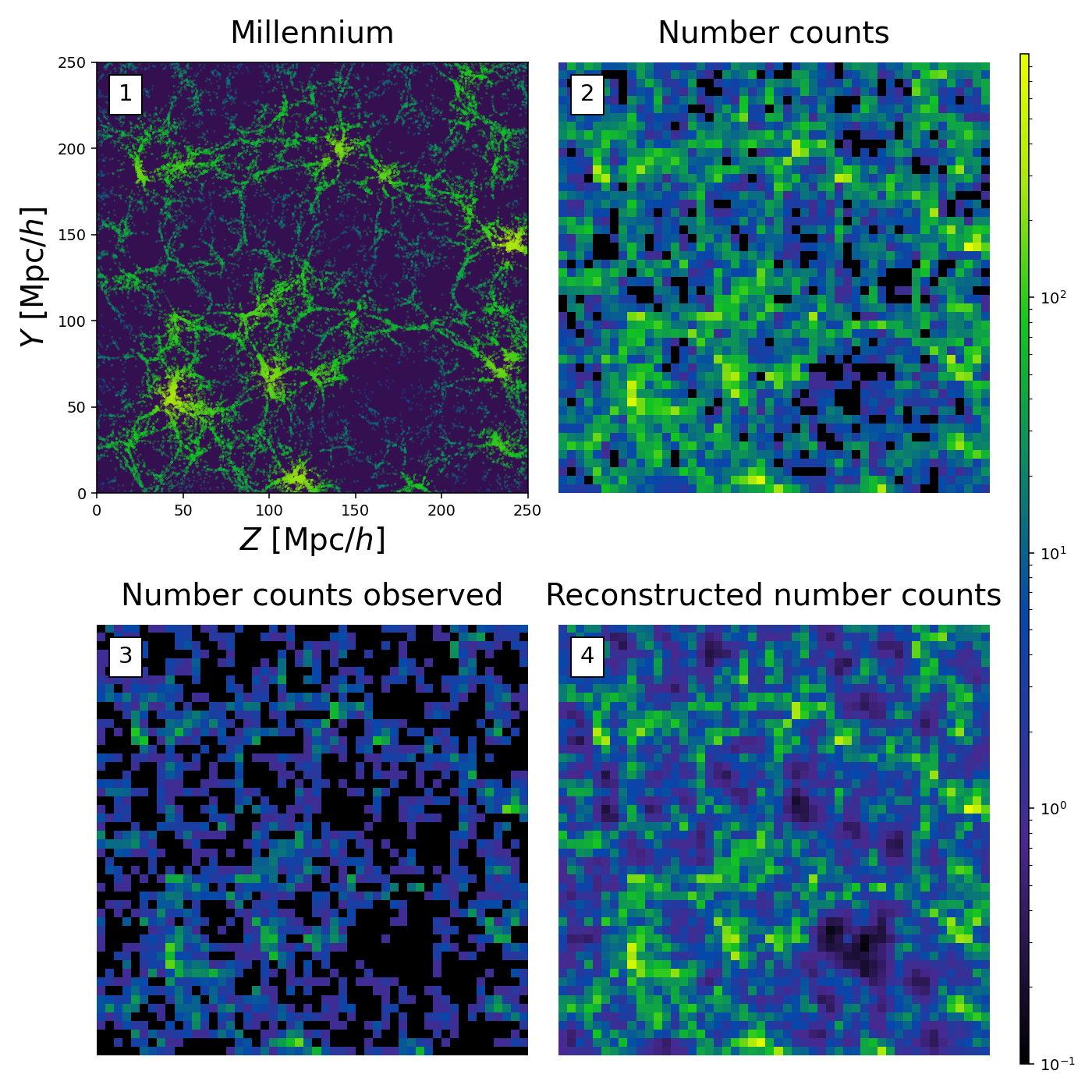}
    \caption{Example of the reconstruction from a $250\,$Mpc$/h$ sub-cube of the Millennium catalog.
    All panels show a slice with the $X$ comoving coordinate held constant. The horizontal axis corresponds to the distance to the observer. 
    Panel (1) shows the individual galaxies in comoving volume, colored according to a Gaussian kernel density estimate (in arbitrary color units). The associated pixelation is drawn in (2), and (3) shows the galaxy number density after imposing an apparent magnitude threshold. Through this magnitude threshold, all number counts are significantly decreased. Panel (4) displays the reconstructed (full) catalog. 
    Our approach succeeds in recovering the original catalog, qualitatively very similar to the true catalog (panel 2). The later Sec.~\ref{subsec: results vanilla case} details the results shown here where we also quantify that the reconstruction is statistically sound. 
    In particular, while panels (2) and (4) do not coincide perfectly, they show reasonable agreement when considering the number count uncertainties not shown here. Consider Fig.~\ref{fig: summary result phenom power spectrum} for the reconstructed galaxy number count uncertainties. }
    \label{fig: example reconstruction}
\end{figure}

As mentioned above, our approach reconstructs the missing galaxy counts following a fully Bayesian approach. As such, we infer the galaxy number count, the parameters of the magnitude distribution, the cosmological parameters and the bias parameters that link the \gls{dm} density contrast with the \gls{dm} density contrast. For each of these quantities, our method produces posterior samples that allow to estimate their mean and uncertainty. 
This comprehensive approach represents a significant advancement over previous methods.
For an example of this reconstruction method, consider Fig.~\ref{fig: example reconstruction}. 

This paper is structured as follows. We begin by summarizing our model for the rate of detected galaxies that relies on a description of the \gls{dm} density contrast. 
To this end, Sec.~\ref{subsec: large scale structure} gives a lightning review of large-scale structure and Sec.~\ref{subsec: matching the 2-point statistics} describes how log-normal fields can be used to model the \gls{dm} density contrast.
The link between the \gls{dm} density contrast to the galaxy rate is subsequently discussed in Sec.~\ref{subsec: bias prescription}. 
To conclude the modeling part, Sec.~\ref{subsec: magnitude introduction} describes the galaxy magnitude distribution that determines the detection probability and, hence, the completeness of the survey, presented in Sec.~\ref{subsec: magnitude distribution and selection effect}.
The method section formulates the likelihood of observing the catalog (Sec.~\ref{subsec: the full likelihood}) and, in particular, we treat how the likelihood incorporates large-scale correlations. 
Sec.~\ref{sec: results} showcases the formalism and its robustness on the example of the Millennium Simulation where we also discuss the implications of redshift uncertainties in Sec.~\ref{subsec: impact redshift uncertainties}. 

A summary of all variables of the analysis can be found in Tab.~\ref{tab: overview variables}. 

\begin{table}[]
    \centering
    \renewcommand{\arraystretch}{1.37}
    \begin{tabular}{cl}
        \hline
        \hline
        \textbf{Variable} & \textbf{Description} \\
        \hline
        \multicolumn{2}{c}{\textbf{Hyperparameters}}\\
        \hline
        $\hyperparams$ & The set of all hyperparameters \\
        $\hyperpowerspectrum$ & Power spectrum parameters \\
        $\hyperGalaxies$ & Bias parameters  \\
        $\hypermagnitudes$ & Parameters of the absolute magnitude distribution \\
        $\hypercosmology$ & Cosmological parameters \\
        \hline
        \multicolumn{2}{c}{\textbf{Cosmological Parameters}}\\
        \hline
        $H_0$ & Hubble constant \\
        $\Omegam$ & Matter density parameter \\
        \hline
        \multicolumn{2}{c}{\textbf{Large-Scale Structure}}\\
        \hline
        $\rhoDM$ & Matter density field \\
        $\rhoDMbar$ & Mean matter density \\
        $\densityDM$ & \gls{dm} density contrast \\
        $\powerspectrumDM$ & \gls{dm} power spectrum \\
        $\rhogbar$ & Mean galaxy density (bias parameter) \\
        \hline
        \multicolumn{2}{c}{\textbf{Galaxy Catalogs}}\\
        \hline
        $\countGalaxiesTrue$ & True galaxy count \\
        $\countGalaxiesObs$ & Observed galaxy count \\
        $M$ & Absolute magnitude \\
        $m$ & Apparent magnitude \\
        $\mthresh$ & Apparent magnitude threshold \\
        $z$ & Redshift \\
        $\skypos$ & Sky position \\
        $\Data$ & Data associated to each galaxy, i.e.~$\Data \deffrom \{z,   \skypos, m \}$ \\
        $\rateGalaxies$ & Mean galaxy rate\\
        \hline
        \multicolumn{2}{c}{\textbf{Field Variables}}\\
        \hline
        $\fieldgausswhitenedspatial$ & Gaussian whitened field in spatial domain \\
        $\fieldgausswhitenedfourier$ & Gaussian whitened field in Fourier domain \\
        $\fieldgausscoloredfourier$ & Gaussian colored field in Fourier domain \\
        $\fieldgausscoloredspatial$ & Gaussian colored field in spatial domain \\
        \hline
        \multicolumn{2}{c}{\textbf{Pixel notation}}\\
        $X_{I}$ & Quantity $X$ in voxel $I$ (e.g.~density) \\
        $X_{\nu, I}$ & $\nu$th quantity $X$ in voxel $I$ (e.g.~individual galaxy magnitudes) \\
        \hline
        \hline
    \end{tabular}
    \caption{Overview of the model variables.}
    \label{tab: overview variables}
\end{table}

\section{Power spectrum and galaxy modeling}

In what follows, we describe the forward modeling of the spatial galaxy distribution. 
Throughout our work, we simplify the geometry of the problem and assume a far-away observer surveying a cubic volume filled with galaxies. 
Each galaxy is characterized by a spectroscopic redshift, a sky position and an apparent magnitude, and, in a first step, none of which have an uncertainty.\footnote{Note that we study the impact of spectroscopic redshift uncertainties on our reconstruction in the later Sec.~\ref{subsec: impact redshift uncertainties}. }
This cube is divided into typically $\mathcal{O}(10^{5})$ voxels. 
We closely follow the prescription for the spatial distribution of galaxies from \cite{Agrawal:2017khv}, assuming a log-normal Gaussian random field to describe the \gls{dm} density, $\densityDM$. The galaxy number rate is then linked to the \gls{dm} density contrast by a local bias prescription. 
Finally, the true number of galaxies is Poisson-sampled from this rate. 

\subsection{Preliminaries}
\label{subsec: preliminaries}

Before we describe the Bayesian approach to the reconstruction, a number of concepts are introduced to model the \gls{dm} density contrast. 

\subsubsection{Large-scale structure and 2-point statistics}
\label{subsec: large scale structure}

We define the \gls{dm} density field as $\rhoDM$, with $\rhoDMbar$ the mean \gls{dm} density $\rhoDMbar \deffrom \langle  \rhoDM\rangle_{V}$, where $\langle  \cdot\rangle_{V}$ stands for average over a given volume. 
From this, we can define the \gls{dm} density contrast in spatial domain \cite{2003moco.book.....D}
\begin{equation}
    \densityDM \deffrom \frac{\rhoDM}{\rhoDMbar} - 1\,.
\end{equation}
Note that $\rhoDM\geq 0$ implies $\densityDM \geq -1$, which will be important for modeling the \gls{dm} density contrast later. 
To characterize the spatial correlation of $\densityDM$ it is common to work in Fourier space. We define the Fourier-transformed field as 
\begin{equation}
    \densityDMFourier(k) \deffrom \int \dd x \,\densityDM(x)\exp\left[-2\pi i x\cdot k\right]\,.
\end{equation}
with $x$ the spatial position vector, in our case, always three-dimensional, and $k$ designates the wave-vector. For clarity, we will denote all variables in Fourier-space with superscripts $(K)$. 
From the Fourier-transformed \gls{dm} density, one can implicitly define the \textit{\gls{dm} power spectrum}
\begin{equation}
\label{eq: def power spectrum}
    \powerspectrumDM(k) \,\deltaDirac(k-k') \deffrom \frac{1}{(2\pi)^3} \langle \densityDMFourier(k)\densityDMFourier(k')^*\rangle_{\samples}
    \,,
\end{equation}
where $\deltaDirac$ is the delta-Dirac distribution, $\langle  \cdot\rangle_{\samples
}$ denotes the mean over sample draws and $*$ is the complex conjugate. 
Since the \gls{dm} density contrast appears twice in Eq.~\eqref{eq: def power spectrum}, this expression is also referred to as the \textit{2-point statistics} of the field. 

\begin{figure}
    \centering
    \includegraphics[width=\textwidth]{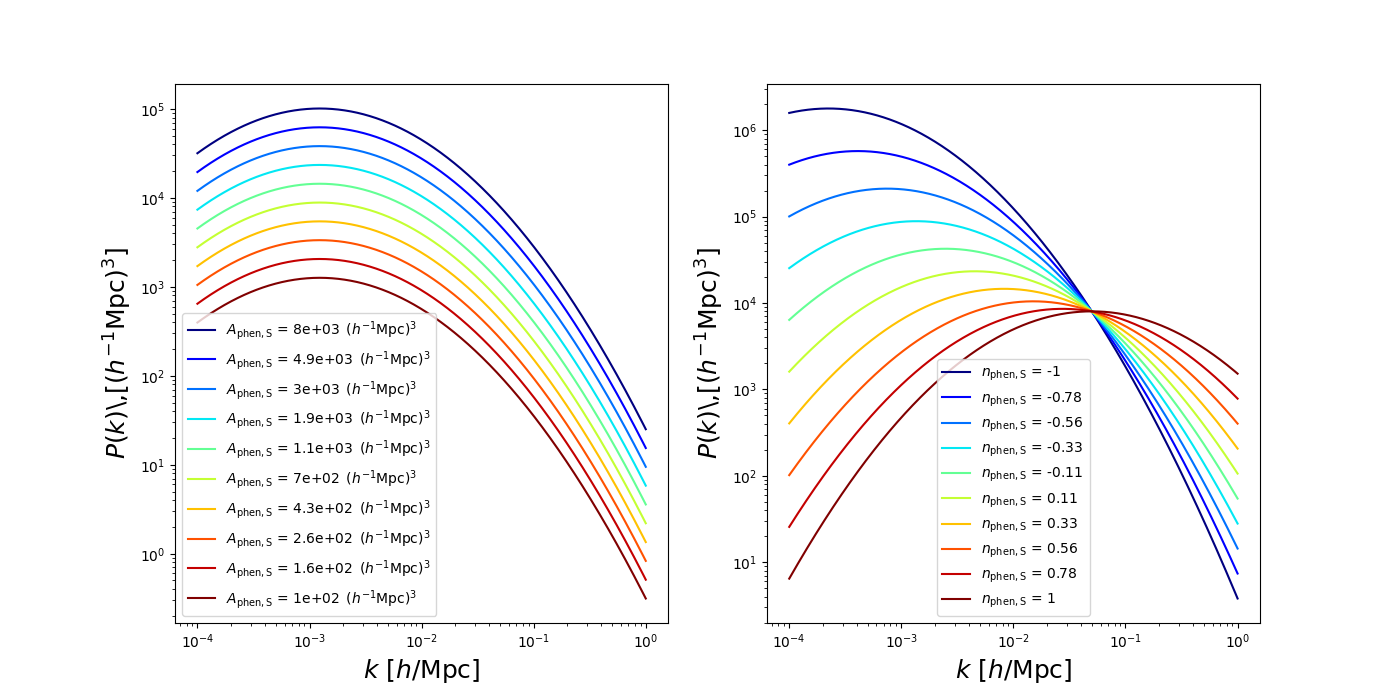}
    \caption{Illustration of the different possible power spectra for varying amplitude $\PSamplitude$ and $\PSn$, as defined in Eq.~\eqref{eq: def phenomenological description power spectrum}. In the left figure, we fix the slope parameter to $\PSn = -1$, whereas the right plot fixes the amplitude to $\PSamplitude = 100\,h^{-1}{\rm  Mpc}$. The remaining parameters are chosen as $\PSalpha=-0.37$ and $k_0=0.05\Mpc/h$.
    }
    \label{fig: illustration power spectrum phenomenological}
\end{figure}

For parts of the results section, we rely on a phenomenological description of the power spectrum, in terms of the parameters $\PSamplitude, \PSalpha,\PSn$ and $k_0$. 
These define the power spectrum via
\begin{equation}
    \label{eq: def phenomenological description power spectrum}
    P(\hyperpowerspectrum;k)
    = \PSamplitude \left( \frac{k}{k_0} \right)^{\PSn - 1 + 0.5 \PSalpha \log\left(\frac{k}{k_0}\right)}
\,.
\end{equation}
Throughout, we fix $\PSalpha=-0.37$ and $k_0=0.05\Mpc/h$, but infer the parameters $\PSamplitude$ and $\PSn$ that govern the overall amplitude and power law decay of the power spectrum. 
For reference, we plot the phenomenological power spectrum for varying values of $\PSamplitude$ and $\PSn$ in Fig.~\ref{fig: illustration power spectrum phenomenological}. 
Since this simplified functional form does not include the effects from non-linear growth at small scales, we will later also use a cosmologically informed power spectrum from \cosmopowerjax{} \cite{SpurioMancini:2021ppk} that accounts for such effects. 

\change{The \cosmopowerjax{} model \cite{Piras:2023aub} is a JAX implementation of several neural emulators \cite{SpurioMancini:2021ppk}. In this work, we use an emulator that approximates the DM power spectrum produced from the Einstein-Boltzmann solver CLASS \cite{Blas:2011rf}. 
The training data are $\sim 5\times 10^5$ CLASS power spectra that include non-linear corrections (computed via \cite{Mead:2015yca, Mead:2016zqy}) due to baryonic effects that depend on the halo concentration $\cmin$ and the halo bloating parameter $\eta_0$. 
This brings the number of input parameters to a total of seven; the two aforementioned parameters in addition to the Hubble constant, $H_0$, the current matter content, $\omegacdm\deffrom \Omega_{\rm cdm} h^2$, and baryonic matter content, $\omegab\deffrom \Omega_{\rm b} h^2$, as well as the initial amplitude and slope of the primordial density fluctuations. The neural network is then trained via minimizing a mean-squared-error loss function. }
We anticipate here that non-linear features of the power spectrum do not noticeably impact the reconstruction of the galaxy number counts (cf.~Sec.~\ref{subsec: comparison power spectrum}).

The above Eq.~\eqref{eq: def power spectrum} between the power spectrum and the average of field realizations assumes that the correlation structure in Fourier space is diagonal. In this case, one can show that the spatial correlation $\xi(x,y)$, defined by 
\begin{equation}
\label{eq: def spatial correlation function}
    \xi(x,y) \deffrom
    \langle \densityDM(x)\densityDM(y)\rangle_{\samples}
\end{equation}
depends on the positions $x$ and $y$ solely through their distance $|x-y|$.
This leads to the short-hand notation for the correlation function as $\xi(z=x-y)\deffrom\xi(x,y)$. 
Note the relation between the power spectrum and the correlation function
\begin{equation}
\label{eq: relation spatial correlation function and power spectrum}
    \xi(x) = \int \dd k \,\powerspectrumDM(k)\exp\left[2\pi i x\cdot k\right]\,.
\end{equation}

\subsubsection{Matching the 2-point statistics}
\label{subsec: matching the 2-point statistics}

As one can theoretically calculate the 1-point and 2-point statistics of the \gls{dm} density contrast, it seems natural to describe this field in terms of a \gls{grf}. 
However, this neglects that the mass density is always positive, i.e.~$\densityDM\geq -1$, which is problematic since large fluctuations of the Gaussian field violate this bound. 
To incorporate this constraint, we follow previous work \cite{Coles:1991if, Agrawal:2017khv} on the generation of simulated galaxy catalogs and assume that an exponentiated \gls{grf}, a log-normal field, to describe the \gls{dm} contrast.\footnote{We discuss the limitations of this approach at the end of this section. }

\begin{figure}
    \centering
    \includegraphics[width=\textwidth]{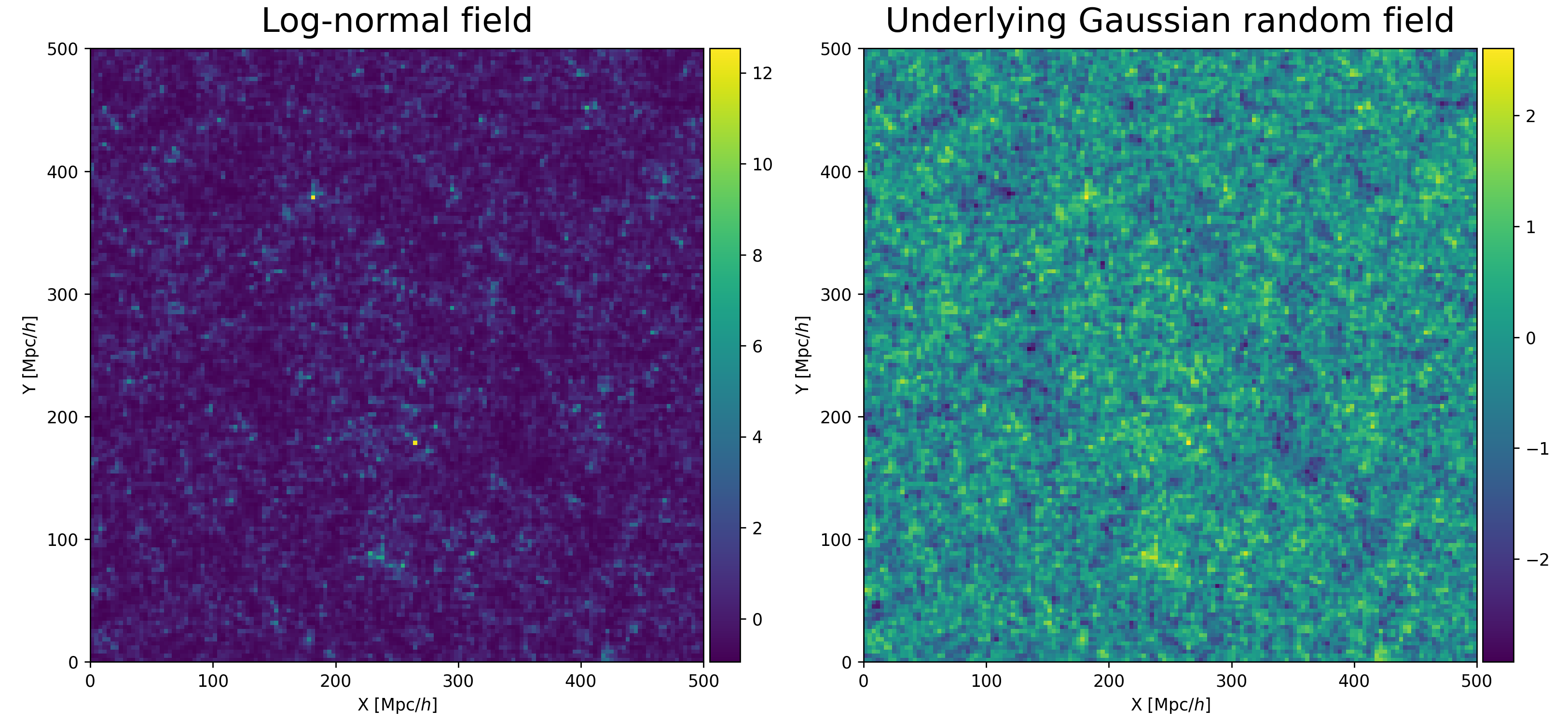}
    \caption{Example two-dimensional slice of a realization of a log-normal field (left) and its associated \gls{grf} (right) in three dimensions. 
    The asymmetry of the log-normal field is clearly apparent, reaching large values in only a few outliers, but (by construction, cf.~Eq.~\eqref{eq: def log-normal field}) never reaching points below $-1$. In this example, we simulate a field of a comoving size of $500$~Mpc$/h$ with 150 bins per dimension. }
    \label{fig: example spatial log-normal field}
\end{figure}

Casting this in mathematical terms, where $\fieldgausscoloredspatial$ denotes the \gls{grf}, we have \cite{Coles:1991if, Agrawal:2017khv}
\begin{equation}
\label{eq: def log-normal field}
    \densityDM \deffrom 
    \exp[\fieldgausscoloredspatial - \sigma ^ 2 / 2]  -1 \,,
\end{equation}
with\footnote{Note that $\sigma$ is defined from the average taken over the ensemble (i.e.~different realizations), not the volume. } $\sigma \deffrom \left[\langle\fieldgausscoloredspatial^2 \rangle_{\samples}\right]^{1/2} $, the (square root of the) variance of the field $\fieldgausscoloredspatial$. This normalization ensures that the mean of the density contrast vanishes. However, this prescription does not enforce that all \textit{realizations} of the \gls{dm} density contrast have mean zero -- this is only true for the mean of the ensemble average. 
See Fig.~\ref{fig: example spatial log-normal field} for one realization of a log-normal field, with its associated \gls{grf}. Compared to the \gls{grf}, the log-normal field allows for higher, positive outliers, and has support over $\left(-1,\infty\right)$, making a suitable description of the \gls{dm} density contrast. 

Using a log-normal field addresses the constraint that the \gls{dm} density is always positive. 
However, whereas it is easy to match the 1-point and 2-point statistics of a \gls{grf}, this is more involved for a log-normal field.
Since we want to model the \gls{dm} density contrast as a log-normal field with a given 2-point statistics, we now present a result that links the statistics of a log-normal field and its associated \gls{grf}. 

Say, we have a \gls{grf}, $\fieldgausscoloredspatial$, with a given power spectrum. The Fourier transform of this power spectrum yields its spatial correlation function $\spatialcorrelationfieldgaussian$. 
Eq.~\eqref{eq: def log-normal field} allows to compute the log-normal field associated to the \gls{grf} and denote the power spectrum of the log-normal field as $P_{\rm LN}^{(K)}$ and its spatial correlation function as $\spatialcorrelationfieldlognormal$. 
Crucially, the two spatial correlation functions are analytically linked via the expression \cite{Coles:1991if}
\begin{equation}
\label{eq: relation spatial correlation function log-normal and gaussian random field}
    \spatialcorrelationfieldgaussian(x) = \log\left[1 + \spatialcorrelationfieldlognormal(x)\right] \,.
\end{equation}
This relation only holds if $1 + \spatialcorrelationfieldlognormal(x) \geq 0$, i.e.~it breaks down if the spatial correlation function attains values below $-1$.
Later, we will choose a target power spectrum for the log-normal field that models the dark matter density field. We can then use the above procedure to obtain the effective power spectrum of the corresponding \gls{grf}.
The above relation is therefore practical for computational reasons since it is straightforward to draw \gls{grf} realizations that then produce a log-normal field with the chosen target power spectrum. 

\begin{figure}
    \centering
    \includegraphics[width=\linewidth]{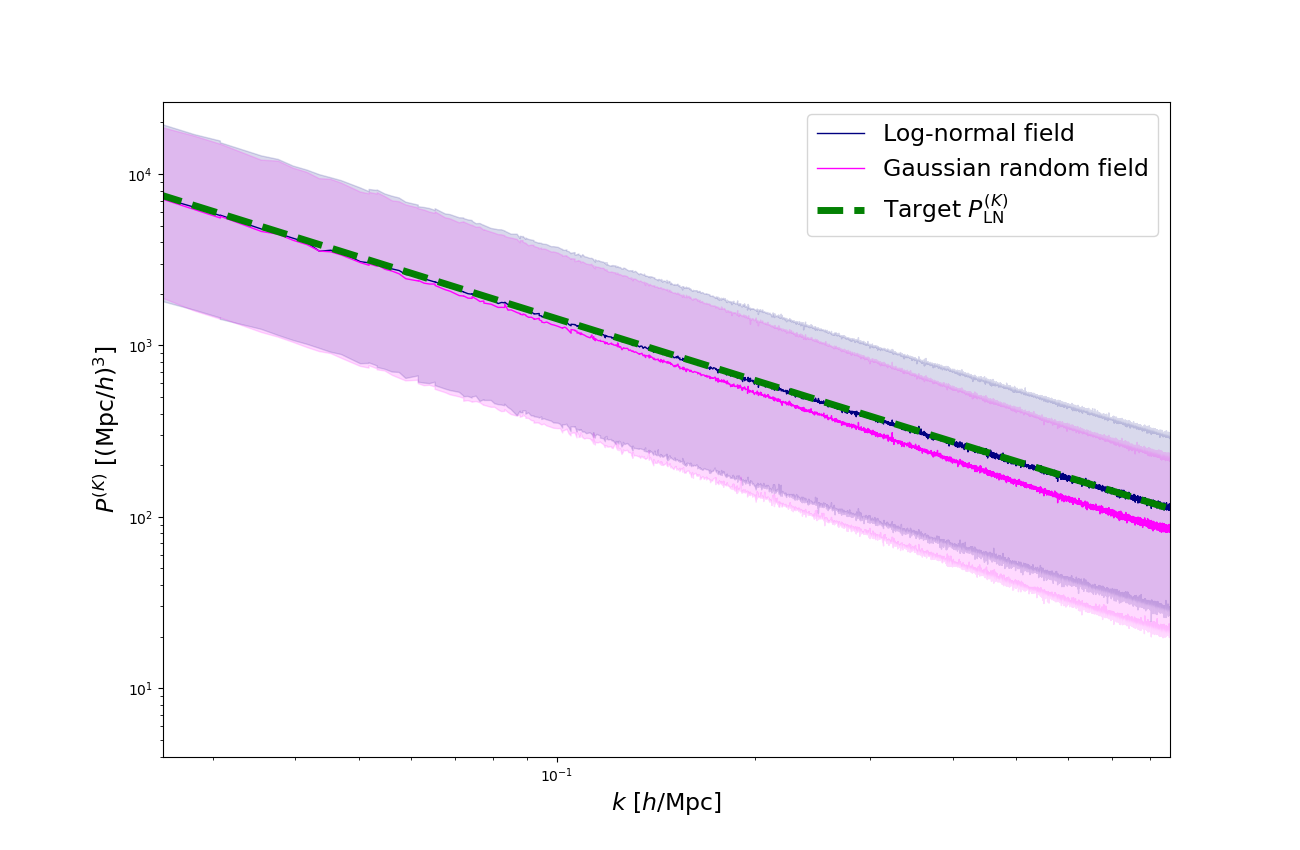}
    \caption{Example for the matching of the 2-point statistics of the \gls{grf} and the log-normal field. 
    We plot the intended power spectrum as reference (green), the measured power spectra of 1000 field realizations of the log-normal field (dark blue) and \gls{grf} (magenta), with the 1~sigma interval as shaded regions around the median (thick dashed line).
    (Fig.~\ref{fig: example spatial log-normal field} shows one realization of the log-normal field and its associated \gls{grf}.)
    The two measured power spectra (dark blue and magenta) differ as expected, but the power spectrum of the log-normal field (dark blue) coincides with the intended power spectra (green), as expected. }
    \label{fig: example power spectrum matching}
\end{figure}

To be as clear as possible, the process to sample log-normal field realizations with a given target spectrum is summarized step by step in App.~\ref{app: generation log-normal field steps}. 
We show an example of this 2-point matching for one realization of the log-normal field in Fig.~\ref{fig: example spatial log-normal field}. 
To verify the 2-point statistics matching, we measure the power spectrum from 1000 realizations of the field in Fig.~\ref{fig: example power spectrum matching} and compare it to the target power spectrum. 
As expected, the power spectrum of the log-normal field coincides with the target power spectrum, but the power spectrum of the log-normal field $\densityDM$ and the Gaussian random field $\fieldgausscoloredspatial$ differ.

Here, we use the log-normal field as an approximation to incorporate spatial correlations in the \gls{dm} density contrast -- a description that is limited. The time-evolved field from an initially Gaussian field at high redshift is not exactly log-normal: it does not describe the correct small-scale statistics of large-scale structure. 
Additionally, the 2-point statistics of a time-evolved field are redshift-dependent and the galaxies are observed on the past light cone, whereas we assume a equal-time \gls{dm} density contrast model, justified by the low redshifts \change{($0.03 \le z\le 0.15$)} we consider.
We acknowledge that this might not be enough when analyzing real data and leave it for future work to account for these effects.

\subsection{Pixelating the catalog and the flat-sky approximation}
\label{subsec: geometry}

We divide the volume of the galaxy survey into finite size voxels and replace all Fourier transforms with discrete Fourier transforms. The set of all voxels $I$ makes up the galaxy survey cube that is denoted as $\mathcal{I}$.

\begin{figure}
    \centering
    \includegraphics[width=0.8\textwidth]{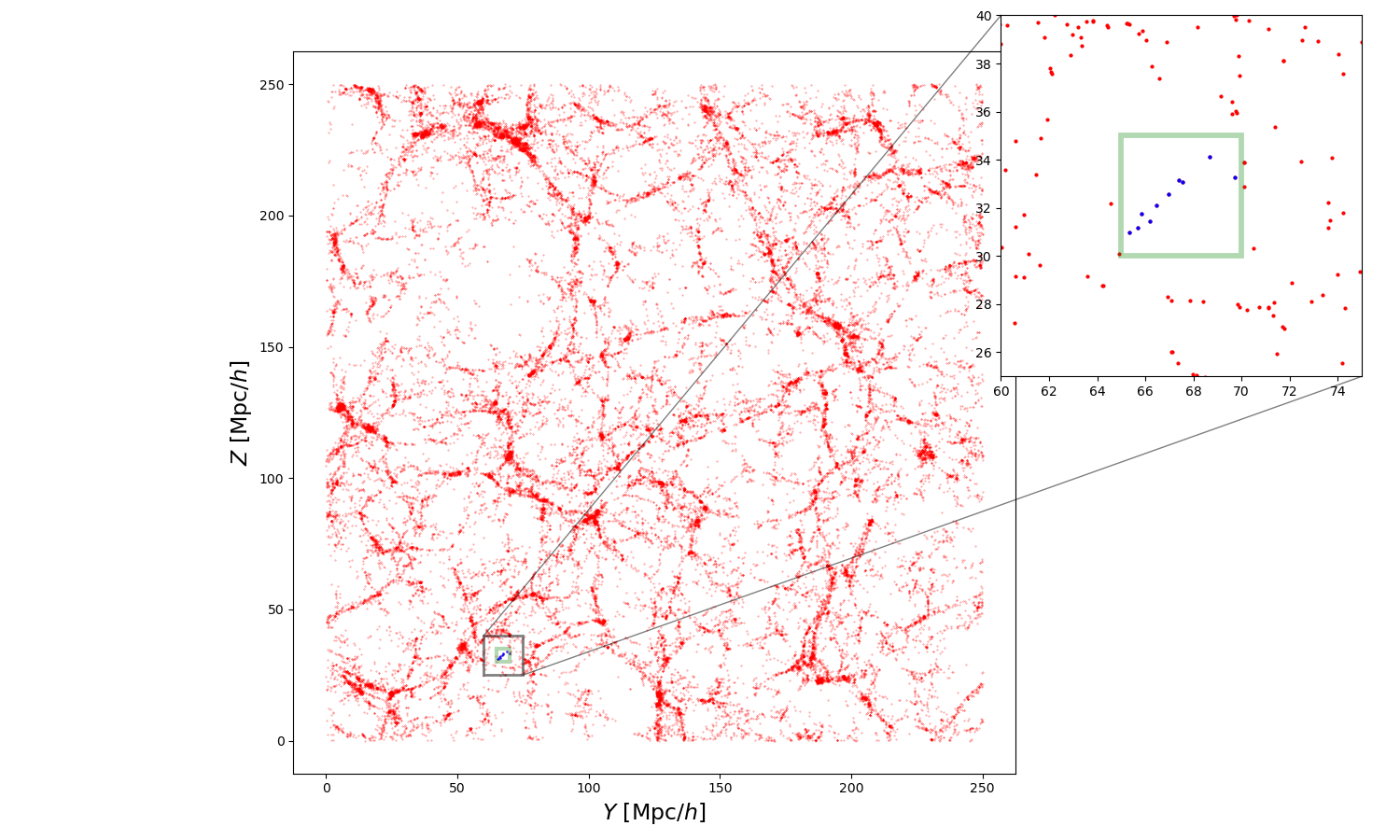}
    \caption{
    Example of a 250~Mpc$/h$ box of the Millennium catalog, with a slice through the $Y-Z$ plane. For this example, we chose 50 segments per side, with an approximate voxel volume of $(5~{\rm Mpc})^3$. This volume varies from voxel to voxel, since the pixelation along the $Z$ axis is uniform in redshift space. The inset shows the outline of one voxel in green, with the galaxies in that voxel drawn in blue. The cube contains a total of approximately 2.9 million galaxies.}
    \label{fig: example pixelization}
\end{figure}

We assume each of the galaxies in the catalog to have an associated data $\Data \deffrom\{z,m,\skypos \}$, with the redshift $ z$, the apparent magnitude $m$ and the sky position $\skypos$.
We assume the \textit{flat-sky} approximation -- the observer is sufficiently far away that the distance to the galaxies is solely determined by their $Z$ comoving coordinate.  
The cube is divided regularly along the transverse direction (perpendicular to the line-of-sight) in \textit{comoving distance} and along the line-of-sight in \textit{redshift}, to obtain the voxels.
We show a two-dimensional slice of the three-dimensional histogram of the mock catalog \cite{Bertone:2007sj} that relies on the Millennium Simulation \cite{Springel:2005nw} in Fig.~\ref{fig: example pixelization}, i.e. a sub-cube of the full simulation with a length of 250~Mpc$/h$, with $h\deffrom H_0 / 100$ the reduced Hubble constant. This sub-cube is divided in 50~segments per dimension, leading to $50^3$ voxels. 

\subsection{Galaxy bias prescription} 
\label{subsec: bias prescription}

In the above description, the \gls{dm} contrast is modeled as a log-normal field. However, numerical simulations indicate that the formation of galaxies is not directly proportional to the \gls{dm} density contrast. 
In general, the number of galaxies that form at a given density of \gls{dm} is a complex process that requires simulations of gravitational interactions, baryonic feedback and hydrodynamics \cite{Kuhlen:2012ft, Reddick:2012qy, Zheng:2015iia, Chaves-Montero:2015iga, Zavala:2019gpq, deMartino:2020gfi, Kokron:2021xgh}.
We take a simplified approach and assume that the \gls{dm} density contrast in each voxel is deterministically related to the rate\footnote{We use the term ``rate'' here as in Poisson rate (expected counts per voxel), not in terms of number counts per units of time. } of galaxies in voxel $I$ via the expression, adapted from \cite{Neyrinck:2013ezr}, given by\footnote{Note that we choose a different parametrization from \cite{Neyrinck:2013ezr}. Rather than fitting the \gls{dm} density below which the galaxy formation is suppressed, $\rho_{\rm exp}$, we introduce $\deltagexp$, the corresponding \gls{dm} density contrast threshold. } 
\begin{equation}
\label{eq: relation density field DM to density field galaxies}
    \rateAbspixel \deffrom \rhogbar \left(1 + \densityDMpixel\right)^\gamma \,
    \exp\left[
        -\left(\frac{1 + \deltagexp}{1 + \densityDMpixel}\right)^{\epsilong}
    \right]
    \, V_{\rm vox}(I)
    \,,
\end{equation}
where $V_{\rm vox}(I)$ is the comoving volume of voxel $I$. 
Note also that we have discretize the \gls{dm} density contrast $\densityDM$ on the same grid as the three-dimensional galaxy voxels, i.e.~$\densityDM\rightarrow\densityDMpixel$.
The variable $\rhogbar$ denotes the mean galaxy density and hence, controls the overall number of galaxies. 
The $\gamma$-term models the non-linear growth of galaxies for large $\densityDM$ with a familiar power law, whereas the exponential term accounts for suppression of the formation of galaxies at low \gls{dm} density, i.e.~for $\densityDM\leq \deltagexp$.
Since the galaxy rate depends solely on the given spacetime point, not its neighborhood, this prescription is called local. 
In the later reconstruction, we will infer all parameters that enter the bias prescription, i.e.~$\rhogbar$, $\gamma$, $\deltagexp$, and $\epsilong$, alongside the missing number counts.
The prior choice of $\gamma, \deltagexp$ and $\epsilong$ does not strongly impact the estimated galaxy number counts, although they correlate with the parameters that determine \gls{dm} power spectrum. 

The above definitions allow one to formulate the expected galaxy mean number rate\footnote{We refer to this variable as a rate since we will assume later that the galaxy number counts are drawn from a Poisson distribution with this term as the rate. } in voxel $I$, $\mu_I(z,\skypos|\hyperparams, \densityDM)$, given the specific realization of the \gls{dm} contrast and the hyperparameters. This density is key to incorporating the large-scale structure information since it propagates the spatial correlation of $\densityDM$ to the distribution of redshift and sky position. 
To formulate the hierarchical likelihood in Sec.~\ref{subsec: the full likelihood}, it will be convenient to factorize the galaxy density 
into an amplitude and spatial distribution as  
\begin{equation}
    \label{eq: rate observation}
     \rateAllpixel{} = \rateAbspixel \,\p(\Data|\hyperparams, \densityDM,I)\,,
\end{equation}
where the overall galaxy rate per voxel, $\rateAbspixel $, was defined in Eq.~\eqref{eq: relation density field DM to density field galaxies} and the distribution $\p(\Data|\hyperparams, \densityDM,I)$ determines the relative frequency with which the data $\Data$ occurs in voxel $I$. 
By definition, we have $ \int \dd\Data\,\rateAllpixel{} = \rateAbspixel$. 
The term $\rateAbspixel$ is crucial for incorporating large-scale structure information into the completed catalog for which we will use the log-normal field formalism described in Sec.~\ref{subsec: matching the 2-point statistics}.
Throughout, we approximate the distribution of galaxies to be uniform in comoving distance $d_c$ and sky position $\skypos$ \textit{within} one voxel, i.e.
\begin{equation}
\label{eq: approximation data likelihood by true parameters}
    \p(\Data=\{z,\skypos\}|\hyperparams, \densityDM,I)
   = 
   \frac{\frac{\dd d_c}{\dd z}}{\Delta d_c(I)\,\Delta \skypos(I)}
   \,,
\end{equation}
where $\Delta d_c(I)$ and $\Delta \skypos(I)$ are the comoving distance and sky position element of voxel $I$.
     
Given the rate of Eq.~\eqref{eq: rate observation}, we assume that the number of galaxies in a redshift bin $z$ and sky position $\skypos$ are Poisson-sampled, i.e.\footnote{See App.~\ref{app: definitions distributions} for a definition of the relevant distributions in this work. }
\begin{equation}
\label{eq: poisson sampling for galaxy rate}
    \p(\countGalaxiesTrue | \Data = \{z, \skypos\},\hyperparams, \densityDM,I) = \Poisson\left(\mu = \rateAllpixel{}; k = \countGalaxiesTrue
    \right) \,,
\end{equation}
where $n_g$ is the number of galaxies in that data bin.
Together, Eq.~\eqref{eq: approximation data likelihood by true parameters} and Eq.~\eqref{eq: poisson sampling for galaxy rate} can be used to define a hierarchical likelihood that describes the probability of drawing $\countGalaxiesTrue$ galaxies with data $\{\Data\}$ in voxel $I$. We detail this likelihood in Sec.~\ref{sec: bayes}.

\subsection{Magnitude modeling and the selection effect}

Galaxy catalogs contain a subset of all galaxies of the Universe -- they are subject to a selection bias. 
While the selection effect is inherently complex for real surveys, this section describes how we approximate it with an apparent magnitude threshold.
The aim of this work is to reconstruct the missing galaxy number counts and therefore, one needs to assess the completeness of the survey. 
To this end, we will also introduce the magnitude distribution of galaxies which, through the apparent magnitude threshold, determines the fraction of detected galaxies and thereby the catalog completeness. 

\subsubsection{Magnitudes and the absolute magnitude distribution}
\label{subsec: magnitude introduction}

All galaxies are characterized by a redshift and a sky position, but in addition also have an absolute magnitude $M$ that is determined by its intrinsic luminosity. 
The rate of Eq.~\eqref{eq: rate observation} is thus modified to 
\begin{equation}
    \mu(z,\skypos|\hyperparams, \densityDM) \rightarrow \mu(z,\skypos,M|\hyperparams, \densityDM)
     = 
     \mu(z,\skypos|\hyperparams, \densityDM)\,\p(M|\hypermagnitudes) \,,
\end{equation}
where the magnitude distribution is defined as
$
    \p(M|\hypermagnitudes)\,,
$
describing the magnitudes of \textit{all} galaxies, both observed and unobserved. 
The set of parameters that defines this distribution is denoted as $\hypermagnitudes$.
The last equality of the above expression relies on the non-trivial assumption that the \gls{dm} density contrast does not impact the magnitude distribution, which will be revised for future work: one expects high-density galaxy clusters to harbor brighter galaxies.  
This approximation will be justified by the verification of our method on simulated data from the Millennium Simulation dataset in the later Sec.~\ref{subsec: results vanilla case}. 
This catalog relies on a full N-body simulation and includes spatial correlation of the magnitudes. 
We will show that one can correctly recover the missing galaxy counts, indicating that the correlation between magnitudes and the \gls{dm} density contrast is not relevant at the current resolution of the reconstruction.
We also model the galaxy magnitudes as independent of redshift, justified by the small redshift ranges \change{($0.03 \le z\le 0.15$)} we consider.\footnote{\change{These redshift ranges result from the distance of the analyzed galaxy cube and its box length, converted with the fiducial value of the cosmological parameters used by the Millennium Simulation, see Sec.~\ref{sec: results} for details. }}
Throughout this work, the magnitude distribution is described by a phenomenological model, specified in App.~\ref{app: magnitude distribution}, see also Fig.~\ref{fig: example magnitude model} for the prior over magnitude distributions. 
The magnitude distribution plays a crucial role as it determines the number of galaxies that remain undetected, as we discuss in the following section. While we are not primarily interested in the precise shape of this distribution, our magnitude distribution incorporates a parameter, denoted as $\muFraction$, which specifies the fraction of galaxies above a fixed magnitude, referred to as $\muThresholdAbove$. 
Later, we will choose $\muThresholdAbove$ to roughly correspond to the faintest absolute magnitude that was detected. 
As this parameter governs the ratio of faint galaxies, it is vital for accurately reconstructing the total galaxy number counts and the impact of its prior will be further discussed in Sec.~\ref{subsec: impact deep survey}.

The apparent magnitude $m$ with which the galaxy is observed can be computed from the absolute magnitude in addition to its luminosity distance $d_L$ with 
\begin{equation}
\label{eq: def apparent magnitude}
    m = M  + 5 \log_{10}\left(\frac{d_L}{1\,\Mpc}\right)
     + 25
    \,.
\end{equation}
The closer the galaxy the \textit{lower} its apparent magnitude. 

\subsubsection{Galaxy survey selection effect}
\label{subsec: magnitude distribution and selection effect}

We assume that the catalog contains a galaxy only if its apparent magnitude (cf.~Eq.~\eqref{eq: def apparent magnitude}) is below a given threshold.
The luminosity distance to a given point in the box, in addition to the apparent magnitude threshold, determines which fraction of the absolute magnitude distribution, $\p(M|\hypermagnitudes)$, is observed. This fraction directly corresponds to the detection probability of each galaxy.
Accordingly, the estimated galaxy number count strongly depends on the inferred magnitude distribution. 
For example, an overestimated fraction of faint galaxies entails overestimating the number of missing galaxies.
Since the magnitude distribution is unknown, we will infer the parameters $\hypermagnitudes$ along with the number of missing galaxies and the other hyperparameters. 

\begin{figure}
    \centering
    \includegraphics[width=\textwidth]{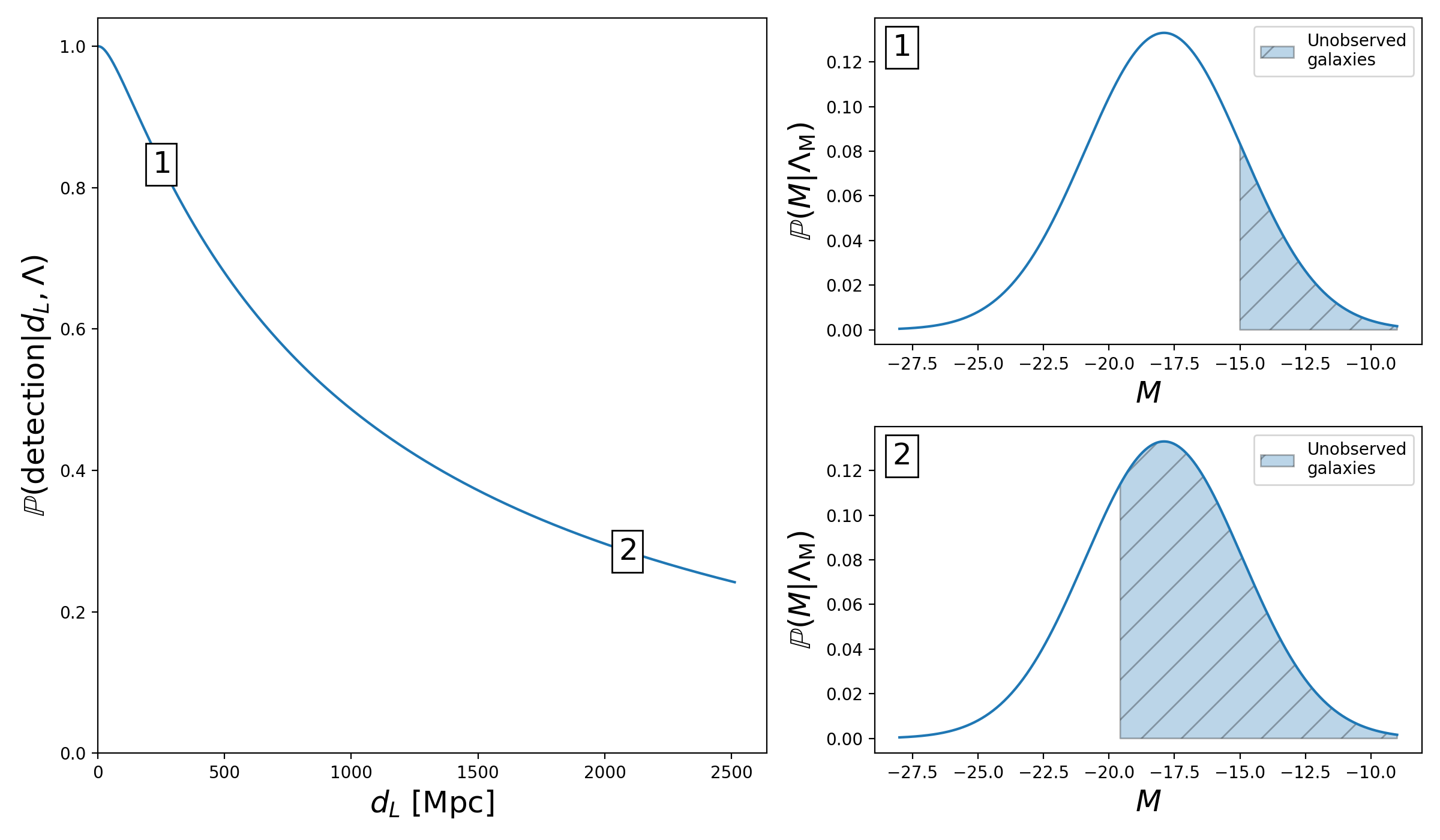}
    \caption{An illustration of the detection probability as a function of the luminosity distance $d_L$ for a Gaussian magnitude distribution, with $ \p(M|\hypermagnitudes) \overset{{\rm here}}{=} \mathcal{N}(\mu=-17.9, \sigma=3; M)$, an approximation to the magnitude distribution of Millennium (the simulated data we consider in later sections).
    We have assumed an apparent magnitude threshold of 22, slightly lower than expected future surveys that attain $\mthresh=25$ \cite{LSST:2008ijt, Euclid:2021icp}. 
    In this simple setup, the detection probability is given by an error function (see Eq.~\eqref{eq: definition detection probability}). 
    For small distances, all galaxies (bright and faint) are detected, leading to high absolute magnitude threshold and a detection probability of one (box 1). With increasing distance, the faint galaxies shift above the apparent magnitude cut, leading to a decreased probability (box 2). 
    For two selected values of the luminosity distance, we plot the corresponding absolute magnitude distributions and their associated observable fraction. }
    \label{fig: illustration detection probability}
\end{figure}

To obtain the probability of \textit{detecting} $\countGalaxiesObs$ out of $\countGalaxiesTrue$ galaxies, one has to marginalize over all undetected galaxies. For a given redshift and cosmological model, this implies an \textit{absolute} magnitude threshold, $\Mthresh(z;\hypercosmology,\mthresh)$, that determines all observable galaxies for this redshift. 
The probability of detection at a redshift $z$ is given by 
\begin{equation}
\label{eq: definition detection probability}
    \p({\rm detection}|z,\hyperparams) = 
    \int ^{\Mthresh(z;\hypercosmology,\mthresh)} _ {-\infty}
    \p(M|\hypermagnitudes)
    \,\dd M
    \,,
\end{equation}
which is the cumulative probability distribution of the magnitude. 
For visualizations, Fig.~\ref{fig: illustration detection probability} shows the detection probability as a function of the luminosity distance $d_L$ on the example of a Gaussian magnitude distribution.
As the luminosity distance increases, the absolute magnitude threshold $\Mthresh(z;\hypercosmology,\mthresh)$ moves to lower values, preventing one from observing fainter galaxies. For small luminosity distances one has $\p({\rm detection}|z,\hyperparams)\approx 1 $, or, the catalog is complete. The detection probability then rapidly descends to lower values, as $\Mthresh(z;\hypercosmology,\mthresh)$ is lowered. Note that the detection probability descends quickest as $\Mthresh(z;\hypercosmology,\mthresh)$ moves through the peak of the absolute magnitude distribution.

\section{Bayesian approach to the reconstruction}
\label{sec: bayes}

Above, we have formulated the model for the \gls{dm} density contrast, its connection to the mean rate of galaxies and the absolute magnitude distribution of galaxies. 
This allows one to assess the probability of observing the data from one galaxy, denoted as $\Data$, given the hyperparameters $\hyperparams$ and a realization of \gls{dm} density contrast $\densityDM$. 
The aim of the following section is to formulate the hierarchical likelihood of an observed galaxy catalog given $\hyperparams$ and $\densityDM$. 
With this likelihood at hand, we can evaluate the posterior distribution which allows us to draw samples in $\hyperparams$ and $\densityDM$ given the observed galaxy catalog. 
Ultimately, from these samples we estimate the galaxy number count in each voxel that represents the reconstructed (i.e.~more complete) galaxy catalog.

\subsection{The hierarchical likelihood}

\label{subsec: the full likelihood}
 
We begin by formulating the likelihood of a galaxy catalog without a magnitude limit that will allow us to formulate the main result: the hierarchical likelihood that links a magnitude-limited galaxy catalog to the hyperparameters and the \gls{dm} density contrast (cf.~Eq.~\eqref{eq: full likelihood marginalized over unobserved galaxy number counts}). 

As an approximation, we assume for a given \gls{dm} density contrast, the galaxy formation in each voxel to be independent. 
Hence, the hierarchical likelihood factorizes to $\p(\Datalist{}|\hyperparams, \densityDM,I)
    =
    \prod_{I\in\mathcal{I}}
    \p(\Datalist{I}|\hyperparams, \densityDM)$, where $\Datalist{I}$ is the data falling in voxel $I$. 
From the assumption that the galaxy data are Poisson distributed (cf.~Eq.~\eqref{eq: poisson sampling for galaxy rate}), App.~\ref{app: derivation hierarchical likelihood} derives the voxel likelihood to be \cite{Mandel:2018mve}
\begin{equation}
\label{eq: rate all galaxies}
    \p(\Datalist{I}|\hyperparams, \densityDM) = 
    \exp\left[
        -  \int \dd\Data \,
    \rateAllpixel{}
    \right]
    \left[
    \frac{1}{{\countGalaxiesTruepixel}!}
    \prod_{\Data\in \Datalist{I}}
    \rateAllpixel{}
    \right]
    \,,
\end{equation}
where we have introduced the voxel-rate in Eq.~\eqref{eq: rate observation}. 

Eq.~\eqref{eq: rate all galaxies} does not account for selection effects.
Since we intend to estimate the galaxy number counts from a magnitude-limited survey, we now derive the analog of Eq.~\eqref{eq: rate all galaxies}, but imposing a selection bias on the data. 
It is computationally involved to sample over all these quantities at once (hyperparameters $\hyperparams$, density contrast $\densityDMpixel$, and the galaxy number counts $\countGalaxiesTruepixel$), i.e.~to draw samples from the posterior according to 
\begin{align}
\label{eq: sampling all parameters illustration}
    \nonumber
    \hyperparams,\densityDM,\countGalaxiesTruepixel &\sim 
    \p(\hyperparams,\densityDM,\countGalaxiesTruepixel|\Datalist{I})
    \\
    &\sim
    \frac{\p(\Datalist{I},\countGalaxiesTruepixel|\hyperparams,\densityDM)
    \,
    \p(\hyperparams,\densityDM)}{\p(\Datalist{I})}
    \,
    ,
\end{align}
with $\p(\hyperparams,\densityDM)$ denoting the prior distribution. 
Therefore, we will derive in the subsequent section the likelihood that is marginalized over the true galaxy number count, i.e.\footnote{Physically, it is unrealistic to have an infinite galaxy count. This results from our assumption that a Poisson distribution governs the galaxy number counts and the Poisson distribution is unbounded. For the numerical implementation we have to choose an upper cut-off for the estimated galaxy number counts above which the posterior has no support. }
\begin{equation}
\label{eq: illustration marginalized likelihood}
    \p(\Datalist{C,I}|\hyperparams,\densityDM)
    =
    \sum_{\countGalaxiesTruepixel=\countGalaxiesObspixel}^{\infty}
    \p(\Datalist{C,I},\countGalaxiesTruepixel|\hyperparams,\densityDM ) \,,
\end{equation}
where the true galaxy number count runs from the observed galaxy number count in voxel $I$, $\countGalaxiesObspixel$, to infinity. 
This marginalized likelihood defines a marginalized posterior from we will draw samples in $\hyperparams$ and $\densityDM$. Only in a second step, we then draw, given the samples in $\hyperparams$ and $\densityDM$, samples in the galaxy number counts via
\begin{equation}
\label{eq: conditional draw galaxy number count}
    \countGalaxiesTruepixel 
    \sim 
    \p(\countGalaxiesTruepixel|\Datalist{I},\hyperparams,\densityDM) = 
    \frac{\p(\Datalist{I},\countGalaxiesTruepixel|\hyperparams,\densityDM)}{\p(\Datalist{I})}
    \,.
\end{equation}

\subsubsection{Likelihood with the true galaxy number count}

We divide the data into sets of detected and non-detected data, respectively, for $\Datalist{} = \bigcup_{I\in\mathcal{I}}\Datalist{C,I} \cup \Datalist{\overline{C},I}$, where the union runs over all voxels $I$ in the galaxy cube $\mathcal{I}$.
The number of elements in each voxel in the detected (non-detected) region is denoted as $\countGalaxiesObspixel$ ($\countGalaxiesNonObspixel$).

We define the rate of detected galaxies in voxel $I$ as 
\begin{equation}
\label{eq: rate det pixel}
    \rateAlldetpixel{}
    \deffrom
    \rateAbspixel \int 
    \dd\Data\,
    \p(\Data|\hyperparams, \densityDM,I)\,
    \p(\text{detection}|\Data,\hyperparams) 
    \,,
\end{equation}
where Eq.~\eqref{eq: definition detection probability}, Eq.~\eqref{eq: relation density field DM to density field galaxies} and Eq.~\eqref{eq: approximation data likelihood by true parameters} can be used to evaluate the above terms. 
The rate of undetected galaxies, $\rateAllnondetpixel{}$, can be defined analogously.
By construction, the relation between the rate of detected and undetected galaxies is
\begin{equation}
\label{eq: sum of rates}
    \rateAllpixel{} = \rateAlldetpixel{} + \rateAllnondetpixel{}\,,
\end{equation}
i.e.~each galaxy falls in one of the two rates of the right hand side. 

Starting from Eq.~\eqref{eq: rate all galaxies} and retaining the number of detected, $\countGalaxiesObspixel$, and undetected galaxies, $\countGalaxiesNonObspixel$, gives\footnote{
Let us stress that this quantity would be much harder to evaluate if the redshift or sky position were uncertain -- in the presence of measurement uncertainties the observed number counts binned by measured values do not coincide with the \textit{observed} number counts binned by the \textit{true} values. Refer to Sec.~\ref{subsec: impact redshift uncertainties} for the impact of spectroscopic redshift uncertainties on the number count reconstruction. }
\begin{align}
    \nonumber
    \p(\Datalist{C,I},\Datalist{\overline{C},I}&,\countGalaxiesObspixel,\countGalaxiesNonObspixel|\hyperparams, \densityDM)
    =
    \exp\left[
        -\rateAbspixel
    \right]
    \frac{1}{(\countGalaxiesObspixel+\countGalaxiesNonObspixel)!}
    {\countGalaxiesObspixel + \countGalaxiesNonObspixel \choose \countGalaxiesObspixel}
    \\
    &
    \left[\prod_{\Data\in \Datalist{C,I}}
    \rateAlldetpixel{}\right]
    \left[\prod_{\Data\in \Datalist{\overline{C},I}}
    \rateAllnondetpixel{}\right] 
    \,.
\end{align}
Let us explain how we have arrived at this expression. The argument of the exponential is the direct result of integration over data of Eq.~\eqref{eq: rate observation}, where we have used $\int \dd \Data \,\p(\Data|\hyperparams,\densityDM,I) =1$. 
We have also replaced the number count $\countGalaxiesTruepixel = \countGalaxiesObspixel + \countGalaxiesNonObspixel$, since each galaxy is either detected or undetected. 
The binomial term is the result of all possible re-orderings of the data points. Finally, depending on whether the galaxy is in the detected or undetected region of data space, it contributes to the detected or undetected galaxy rate, $\rateAlldetpixel{}$ and $\rateAllnondetpixel{}$, respectively.

The undetected data points are by definition unrecognized and their parameters are unknown. Thus, we marginalize over these undetected data but retain the missing galaxy number count, $\countGalaxiesNonObspixel$, which gives using Eq.~\eqref{eq: rate det pixel}
\begin{align}
\label{eq: likelihood data with ng depedency}
    \p(\Datalist{C,I},\countGalaxiesObspixel,\countGalaxiesNonObspixel|\hyperparams,\densityDM ) 
    &
    = 
    \exp\left[
        - 
        \rateAbspixel
    \right]
    \frac{1}{(\countGalaxiesObspixel+\countGalaxiesNonObspixel)!}
    {\countGalaxiesObspixel+\countGalaxiesNonObspixel \choose \countGalaxiesObspixel}
        \\
    \nonumber
    &
    \Biggl[\prod_{\Data\in\Datalist{C,I}}
    \p(\Data|\hyperparams,\densityDM,I)\Biggr]
    \rateAbspixel^{\countGalaxiesObspixel+\countGalaxiesNonObspixel}
    \left[
    \pnotdetpixel
    \right] ^ {\countGalaxiesNonObspixel}
    \,,
\end{align}
where we have defined the detection probability to land in voxel $I$ as $\pdetpixel  \deffrom \frac{\rateAlldetpixel{} }{ \rateAbspixel}$ and accordingly, the non-detection probability as $\pnotdetpixel = 1- \pdetpixel$. 
Recall that the probability distribution $\p(\Data|\hyperparams,\densityDM, I)$ appearing in Eq.~\ref{eq: likelihood data with ng depedency} was defined in Eq.~\eqref{eq: approximation data likelihood by true parameters}. 
The above expression is the likelihood of observing a magnitude-limited catalog, with a galaxy number count of $\countGalaxiesTruepixel=\countGalaxiesObspixel+\countGalaxiesNonObspixel$, given $\hyperparams$ and $\densityDM$. This expression is difficult to sample from, as explained in the following section. 
For a discussion of the maximum likelihood estimator for the galaxy number count, see App.~\ref{app: derivation maximum likelihood estimator for the true number count}. 


\subsubsection{Likelihood marginalized over the galaxy number count}

With the additional assumption of a prior, Eq.~\eqref{eq: likelihood data with ng depedency} defines the posterior distribution appearing on the right-hand side of Eq.~\eqref{eq: sampling all parameters illustration} and, hence, allows one to produce samples of the hyperparameters, the dark matter contrast and the galaxy number counts from a magnitude-limited galaxy catalog. 
However, producing samples from a high-dimensional posterior (for more parameters) is considerably more challenging than sampling from a low-dimensional posterior.
For this reason, it is easier to first produce posterior samples in the hyperparameters $\hyperparams$, and the \gls{dm} density contrast, $\densityDMpixelall$, and then in a second step to draw samples in the galaxy number counts $\countGalaxiesTruepixelall$, as we have discussed in the introduction of Sec.~\ref{sec: bayes}. 
Therefore, we now derive the likelihood but marginalized over the galaxy number count. 

This marginalized likelihood (cf.~Eq.~\eqref{eq: illustration marginalized likelihood}) can be obtained directly from Eq.~\eqref{eq: likelihood data with ng depedency} using $\sum_{\nu=0}^{\infty}\frac{x^{\nu}}{\nu!} = e^x$, leading to
\begin{align}
\label{eq: likelihood observations from hyperparameters}
\nonumber
    \p(\Datalist{C,I}&,\countGalaxiesObspixel|\hyperparams,\densityDM)
    =
    \\
    &
    \exp\left[
        -
        \int\dd \Data \,
    \rateAlldetpixel{}
    \right]
    \left[
     \frac{1}{\countGalaxiesObspixel!}
    \prod_{\Data\in \Datalist{C,I}}
    \left[
    \rateAbspixel
   \,\p(\Data|\hyperparams,\densityDM,I)
    \right]
    \right]
    \,,
 \end{align}
where the relation $\rateAlldetpixel{} = \rateAll{} - \rateAllnondetpixel{}$ (cf.~Eq.~\eqref{eq: sum of rates}) was used. 
Also note that the above holds since $\p(\text{detection}|\Data, \hyperparams,I)=1$ by definition of the data being included in the observations. 
Employing Eq.~\eqref{eq: rate det pixel} and Eq.~\eqref{eq: approximation data likelihood by true parameters}, the above expression reduces to 
\begin{align}
\label{eq: full likelihood marginalized over unobserved galaxy number counts}
\nonumber
    \p(\Datalist{C,I}&,\countGalaxiesObspixel|\hyperparams,\densityDM)
    =
    \\
    \nonumber
    &
    \exp\left[
        -
        \rateAbspixel\int\dd z\,\dd\skypos\,\dd M \,
    \p(z,\skypos,M,\text{detection} |\hyperparams, \densityDM,I){}
    \right]
     \\
    &
     \left[
    \frac{\rateAbspixel^{\countGalaxiesObspixel}}{\countGalaxiesObspixel!}
    \prod_{\Data\in\Datalist{C,I}}
   \p(z(\Data),\skypos(\Data)|\hyperparams, \densityDM,I)\,\p(M(\Data)|\hyperparams)
    \right]
    \,.
\end{align}
Given a magnitude-limited survey, we use the product over all voxel of Eq.~\eqref{eq: full likelihood marginalized over unobserved galaxy number counts} to draw posterior samples in $\hyperparams$ and the \gls{dm} density contrast in each voxel which is simpler since it marginalizes over all possible galaxy number counts. Even so, we will perform parameter estimation for about $\mathcal{O}(10^{5})$ parameters requiring advanced sampling techniques (described in the next section).  
To also obtain posterior samples in the galaxy number count, we can then sample from Eq.~\eqref{eq: likelihood data with ng depedency} with the posterior samples in $\hyperparams$ and $\densityDM$ previously computed (cf.~Eq.~\eqref{eq: conditional draw galaxy number count}).

\subsection{Approximating the posterior with Hamiltonian Monte-Carlo}
\label{subsec: hmc and posterior sampling}

\begin{figure}[htbp]
  \centering
  \resizebox{\textwidth}{!}{%
\input{schemata}
}
  \caption{Overview of the approach to the galaxy number count reconstruction. The hyperparameters including cosmological parameters, $\hypercosmology$, the magnitude model parameters, $\hypermagnitudes$, and the galaxy bias parameters, $\hyperGalaxies$, are denoted as $\hyperparams$.
  As a first step, we produce a \gls{dm} density contrast from a Gaussian random field realization, $\fieldgausswhitenedspatial$, to generate a log-normal field. The 2-point statistics of this log-normal field are then matched to the power spectrum defined by $\hyperpowerspectrum$ and $\hypercosmology$. 
  From this, we compute the mean galaxy rate via the bias prescription (cf.~Eq.~\ref{eq: relation density field DM to density field galaxies}).
  Through the magnitude distribution defined by parameters $\hypermagnitudes$, the rate of detected galaxies is obtained. Finally, this quantity is compared to the observed data.
  The posterior samples of true number counts are obtained a posteriori, after having produced posterior samples in $\hyperparams$ and $\densityDM$. 
  }
  \label{fig: overview}
\end{figure}

Since we estimate the \gls{dm} density contrast for each voxel, the posterior is high-dimensional and we rely on Hamiltonian Monte-Carlo to produce samples from the posterior. 
We use the No-U-turn sampler (NUTS) \cite{Neal:2011mrf, Hoffman:2011ukg, Betancourt:2017ebh} that is implemented in the \numpyro{} package \cite{Phan:2019elc}.\footnote{We have also tested the custom Gibbs sampler developed in \cite{coleman_krawczyk_2024_12167630}, but find no improvement over the analysis using no Gibb's sampling in this particular case. }
The package \numpyro{} relies on JAX to keep track of all Jacobians, relating the whitened \gls{grf} to the density of galaxies $\rateGalaxies$. This is important, since the galaxy rate is deterministically related via the 2-point statistics matching and the bias prescription of Eq.~\eqref{eq: relation density field DM to density field galaxies} to the whitened Gaussian field. Sampling in the whitened Gaussian field considerably facilitates the sampling process since the geometry of the posterior is simpler.
For an overview of our method, consult Fig.~\ref{fig: overview}. 

The sampling includes a warm-up process of 2000 steps during which the mass matrix is tuned, followed by producing 3000 posteriors samples per chain, with four chains in total. 
Samples that result from a trajectory that find no U-turn (after a predetermined number of steps) are labeled as ``divergent''. 
Since this indicates that the parameter space is not properly explored we discard all chains with a relative fraction of divergent samples above 2$\%$. 
During the sampling process the step size is modified with the so-called target acceptance probability. For higher values, the step size become smaller, translating into a slower sampling process but more reliable results. For our setup, we choose a target acceptance probability of 0.9. 
For the prior ranges considered, we refer the reader to App.~\ref{app: prior range}. 
We verify the convergence of the chains by $\hat r \leq 1.05$, the Gelman-Rubin criterion \cite{Gelman:1992zz} for all variables. Occasionally some of the $\hypermagnitudes$ parameters depart from this criterion since different chains find different $\hypermagnitudes$. 
This is not surprising, distinct $\hypermagnitudes$ can lead to very similar magnitude distributions and hence the reconstruction will be virtually identical.  
This is therefore not concerning since their precise values of the $\hypermagnitudes$ are not relevant to the reconstruction;---these parameters are treated as marginal parameters. 

To accelerate the posterior computation, we rely on the use of \glspl{gpu}. 
The typical computation time for one \change{galaxy catalog reconstruction and the production of 3000 associated posterior samples per chain} is $6-22$h on an A100 GPU.

\section{Results}
\label{sec: results}

\subsection{Validation through the Millennium Simulation}

In order to verify our method on realistic data, we use the mock galaxy catalogs of \cite{Bertone:2007sj} that rely on the Millennium Simulation \gls{dm} density field \cite{Springel:2005nw}. 
The catalog consists of a cube of length $500\Mpc/h$ in comoving distance.
Out of this larger box, we choose two smaller sub-cubes. 
For Sections~\ref{subsec: results vanilla case},~\ref{subsec: impact redshift uncertainties} and \ref{subsec: impact deep survey} we use a galaxy box of 250~Mpc$/h$ with an apparent magnitude threshold of 17, resulting in 450,563 observed galaxies out of 2,946,581. 
For Section~\ref{subsec: comparison power spectrum}, we analyze a slightly smaller 150~Mpc$/h$ sub-cube, and impose an apparent magnitude threshold of 21, leading to 348,944 observed galaxies, from originally 589,144 galaxies.
Tab.~\ref{tab: summary result sections} summarizes the different setups and their motivation for a better overview. 
We choose to not analyze larger catalogs, since this would significantly increase the computation time. 
We divide the respective volumes in three-dimensional pixels (voxels)\footnote{The catalog is binned regularly in comoving distance in the $X$ and $Y$ directions that are orthogonal to the line-of-sight and regularly in redshift along the line-of-sight. We always assume the same number of bins in all three directions. For a visualization of the binning of the catalog see Figure~\ref{fig: example pixelization}.  }, as described in Sec.~\ref{subsec: geometry}. The K-band luminosities are provided in the catalog. 
For the data simulation process (not during inference), we assume a flat $\Lambda$CDM cosmology, with $H_0 = 73\,\hu$ and $\Omegam = 0.25$, the values used for the Millennium Simulation. 
Using the flat-sky approximation, we determine the distance to all galaxies solely from their $Z$ comoving coordinate. 
With the assumed cosmological parameters, we convert the comoving distance of each galaxy to a redshift and a luminosity distance. Finally, the apparent magnitude is obtained from the galaxy's luminosity distance and its K-band absolute magnitude.

Below, we test the various assumptions of our approach, starting with the impact of the assumed power spectrum on the reconstructed galaxy counts.
Since we assume perfect spectroscopic measurements of the galaxies' redshift, we assess in the second part the bias induced when galaxy redshifts are imperfectly measured, an uncertainty included in the data simulation part, but neglected during the reconstruction. 
Finally, we present how a deeper survey that is restricted to $\sim1\%$ of the sky strongly improves the reconstruction, both of the galaxy number counts and of the magnitude distribution. Tab.~\ref{tab: summary result sections} summarizes the purpose and assumptions of all following four sections.

\begin{table}[]
\centering
\renewcommand{\arraystretch}{1.9} 
\begin{tabular}{|c|c|c|c|c|}
\hline
\textbf{} & \multicolumn{4}{c|}{\textbf{Sections}} \\ 
\hline
\textbf{} & \textbf{4.2} & \textbf{4.3} & \textbf{4.4} & \textbf{4.5} \\
\hline
\makecell{\textbf{Purpose} \\ \textbf{of section}} & \makecell{Demonstration \\ of method} & \makecell{Impact of \\ power spectrum} & \makecell{Impact of \\ uncertain redshifts} & \makecell{Impact of deeper\\ survey for 25/2500 \\ sky pixels \\ (with $\mthresh=24$)} \\
\hline
\makecell{\textbf{Box Size} \\ (Mpc/$h$)} & 250 & 150 & 250 & 250 \\
\hline
\makecell{\textbf{Redshift} \\ \textbf{Errors}} & No & No & Yes & No \\
\hline
\makecell{$m$ \\ \textbf{thresh.}} & 17 & 21 & 17 & 17 \\

\hline
\makecell{\textbf{Segments} \\ \textbf{per dim.}} & 50 & 40 & 50 & 50 \\
\hline
\makecell{\textbf{Distance}\\ \textbf{to observer} \\ (Mpc$/h$)} & 300 & 250 & 300 & 300 \\
\hline
\end{tabular}
\caption{Summary of result sections, with their assumed survey properties and test purposes.}
\label{tab: summary result sections}
\end{table}

\subsection{Reconstruction with a phenomenological power spectrum model}
\label{subsec: results vanilla case}

\begin{figure}
    \centering
    \includegraphics[width=\linewidth]{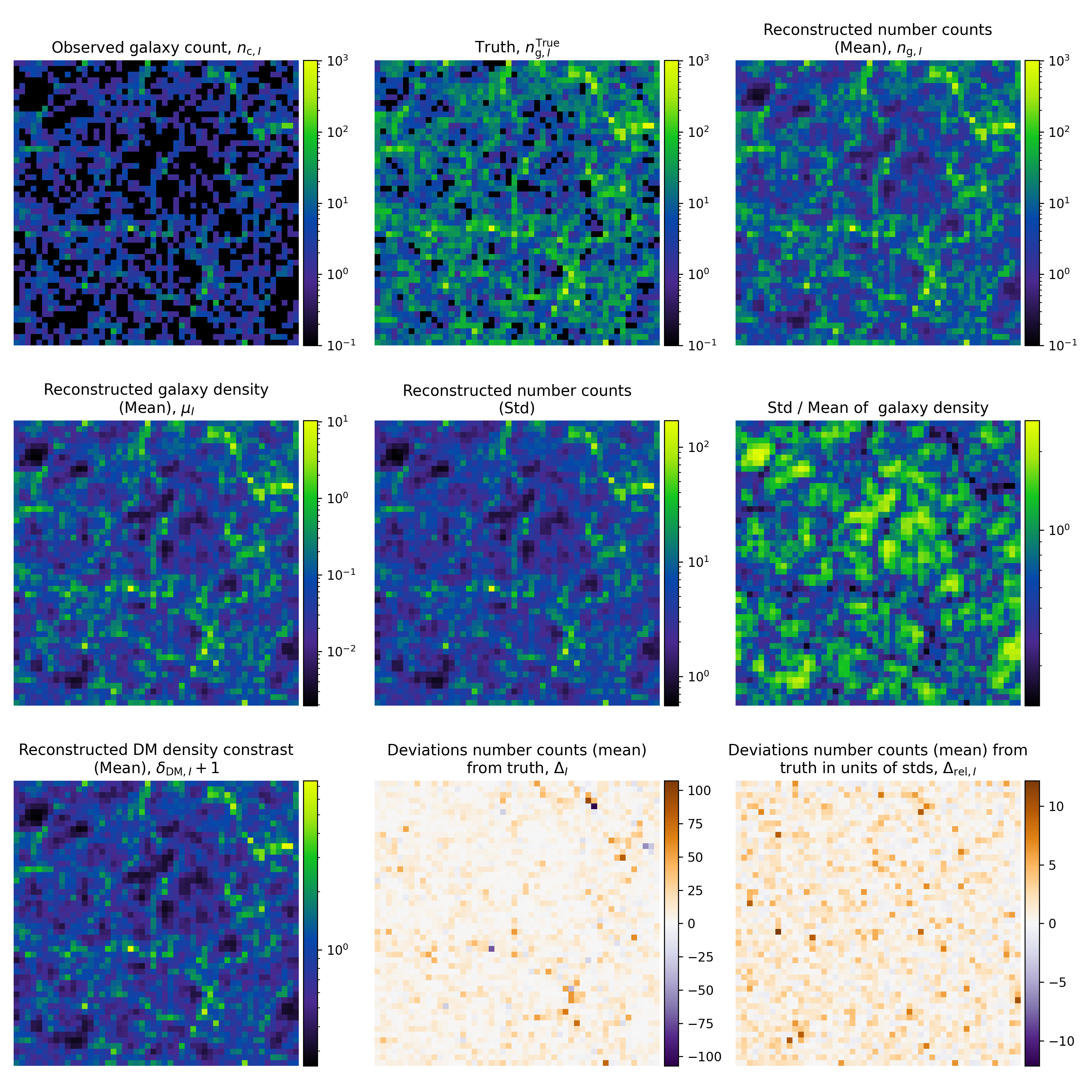}
    \caption{ 
    A slice perpendicular to the $X$-axis of the reconstructed catalog using a phenomenological power spectrum. 
    The observer is located at the left, which (under the flat-sky approximation) gives the horizontal axis the interpretation of the redshift (or distance) coordinate. The box length is 250~Mpc$/h$. 
    From top to bottom, left to right: (1) Observed galaxy count: the galaxy number counts after imposing the apparent magnitude threshold, used as input data. (2) The true number count from which panel (1) is obtained and that is to be compared to the reconstruction in (3): the mean of the reconstructed galaxy number count. (4) The mean of the reconstructed galaxy density $\rateAbspixel$ (cf.~Eq.~\eqref{eq: relation density field DM to density field galaxies}) from which we Poisson-sample the number counts. (5) The standard deviation of the reconstructed galaxy number counts. (6) The ratio between standard deviation and mean of the galaxy density, high density regions have lower relative uncertainty than low-density regions. (7) The mean of the reconstructed \gls{dm} density contrast ($+1$). (8) The absolute deviations between truth and reconstructed number counts (cf.~Eq.~\eqref{eq: def absolute deviation}). (9) The relative deviations between truth and reconstructed number counts (cf.~Eq.~\eqref{eq: def relative deviation}).
    Note that the absolute deviations (8) are centered around zero, but regions of large galaxy density have negative deviations. This means that our reconstruction overestimates the number of missing galaxies. Furthermore, the relative deviations are strongly asymmetric, the reason for which is explained in the main text.}
    \label{fig: summary result phenom power spectrum}
\end{figure}

As a first step, we assume a phenomenological power spectrum determining the 2-point statistics of the log-normal field that was defined in Eq.~\eqref{eq: def phenomenological description power spectrum} to govern the power spectrum of the \gls{dm} density contrast.

\begin{figure}
    \centering
    \includegraphics[width=\textwidth]{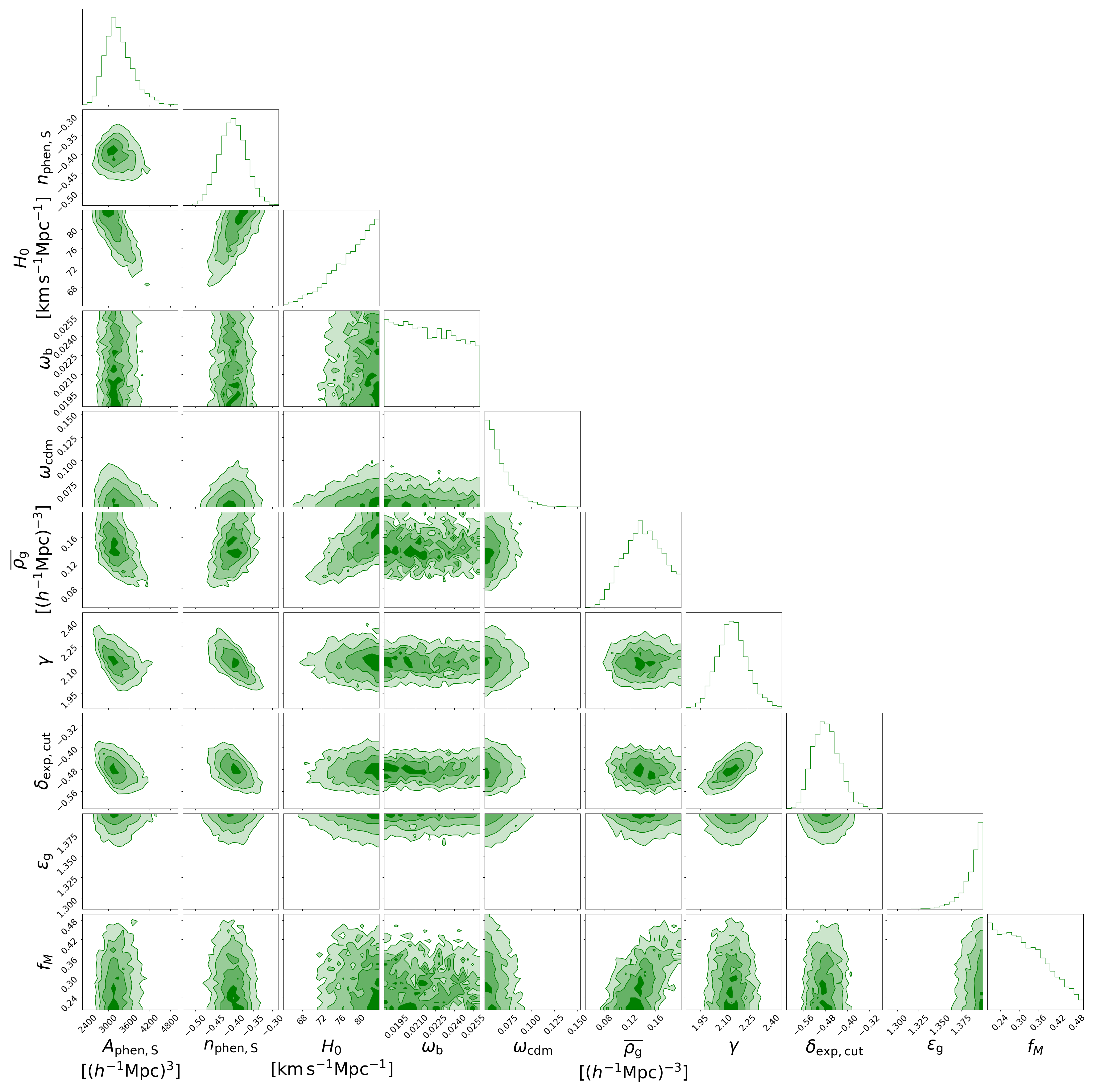}
    \caption{A corner plot, showing a subset of the two-dimensional marginal hyperparameter posteriors. The Hubble constant correlates strongly with the overall amplitude of the power spectrum $\PSamplitude$ since both variables govern the overall number of galaxies in each voxel. The Hubble constant impacts the galaxy number count since it links it to the galaxy density via the voxel volume (cf.~Eq.~\eqref{eq: relation density field DM to density field galaxies} where the voxel volume has the approximative  $1/H_0^3$ dependency in the flat-sky approximation). The power spectrum parameter $\PSamplitude$ also determines the galaxy number count via the \gls{dm} density contrast and Eq.~\eqref{eq: relation density field DM to density field galaxies}. The same reasoning can be applied to the mean density of galaxies $\rhogbar$, which controls the expected galaxy number count. 
    \change{The above constraint on $H_0$ is thus driven by volume effects and would change if different conditional priors on $\PSamplitude$ or $\rhogbar$ were assumed. 
    This highlights the importance of prior assumptions on the reconstructed catalog and the value of the inferred cosmological parameters. }
    }
    \label{fig: corner plot vanilla case}
\end{figure}

\begin{figure}
    \centering
    \includegraphics[width=0.9\textwidth]{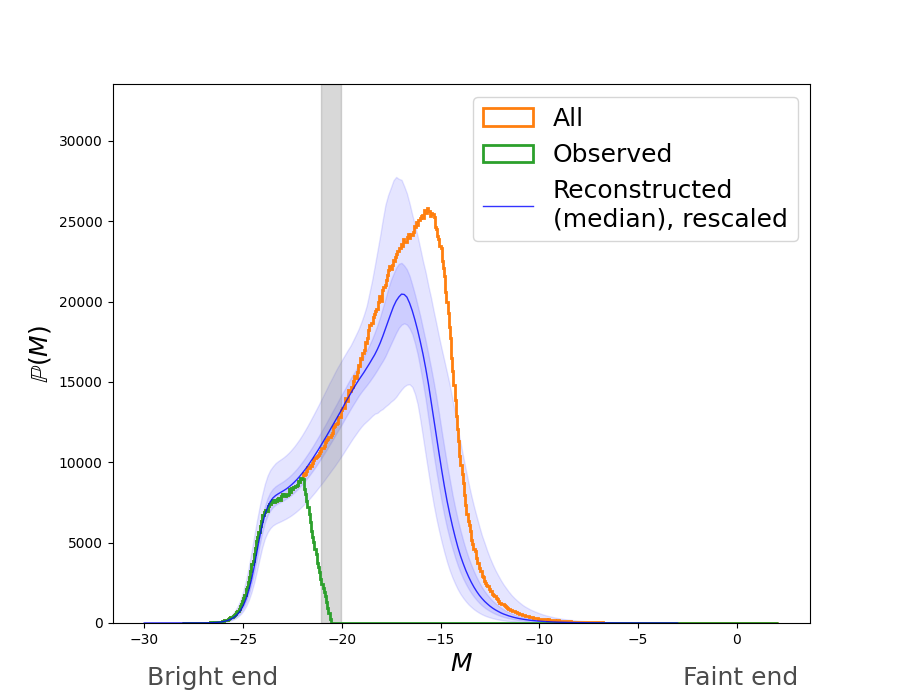}
    \caption{The magnitude distribution, where the orange (green) line shows the distribution of all (detected) galaxies.
    As expected, the two distributions coincide at lower magnitude and differ strongly at the faint end (high magnitudes). We have indicated the faintest galaxy that is observed with a grey band -- above this point, the data provides no constraint on the distribution of magnitudes and the results are no longer trustworthy. 
    The reconstruction (with the median in blue) is not accurate at the faint end: the inferred magnitude distribution is smaller than the true underlying population of faint galaxies, and therefore, the number counts of faint galaxies are expected to be underestimated. For easier visual comparison, we have rescaled the magnitude distribution to agree with the histogram of detected galaxies at low magnitudes. 
    Finally, the blue envelopes indicate the 1 and 2~sigma intervals of the reconstructed magnitude distribution.  
    The uncertainty bands are wider at the faint end, since fewer samples provide constraints in this parameter region. }
    \label{fig: magnitudes reconstructed}
\end{figure}

Following the procedure using Hamiltonian Monte-Carlo for sampling (see Sec.~\ref{subsec: hmc and posterior sampling}) with \numpyro{}, with the likelihood of Eq.~\eqref{eq: full likelihood marginalized over unobserved galaxy number counts}, we obtain the result that is summarized in Fig.~\ref{fig: summary result phenom power spectrum}.
We show a two-dimensional slice (of the inherently three-dimensional recovered galaxy number counts), holding the $X-$dimension of the catalog constant. 
The reconstruction marginalizes over the bias parameters, cosmological parameters and magnitude parameters, and we plot the two-dimensional marginal distributions for selected parameters in Fig.~\ref{fig: corner plot vanilla case}. 
Since the magnitude distribution determines the catalog completeness, it strongly impacts the reconstructed number counts. The inferred magnitude distribution and its uncertainty are shown in Fig.~\ref{fig: magnitudes reconstructed}, with the detected absolute magnitudes and all absolute magnitudes histogrammed for reference. 
The bright end of the magnitude distribution is faithfully recovered, and the overall spread of the distribution is small since there are many observed galaxies. 
However, for faint galaxies, the inferred and true magnitude distributions differ more and more. This discrepancy is the result from the lack of constraining power of the data for magnitudes above the faintest galaxy that is included in the catalog (grey shaded band). Above that value, the inferred magnitude distribution is extrapolated and is prior dependent. 
We also note the importance of the parameters governing the fraction of unseen galaxies, $\muFraction$, at a reference absolute magnitude, $\muThresholdAbove$. The magnitude distribution will not be recovered correctly if one fixes $\muFraction$ to an incorrect value.
Here, and for all other results shown, we marginalize over this parameter. 
Even so, although the uncertainty of the reconstruction increases for faint magnitudes, it does not fully encompass the true magnitude distribution. This failure indicates that subsequent work has to consider a more agnostic parametrization of the magnitude distribution.

For \gls{gw} cosmology with current detector sensitivities, the incorrect recovery of the faint magnitude distribution is not necessarily a reason for concern since faint galaxies are much less likely to host \gls{gw} sources. Indeed, the luminosity approximates the stellar mass of the galaxy and hence, the rate with which it produces \glspl{bbh}. Consequentially, these faint galaxies contribute insignificantly to the redshift distribution of \gls{bbh} and the $H_0$ measurement is not impacted strongly by their abundance, at least at current sensitivity \cite{LIGOScientific:2021aug}.\footnote{Note that the choice of luminosity weighting might be relevant for future studies that use more complete galaxy catalogs. } \change{To cast this reweighting in mathematical terms, one can introduce $p(\text{host})$, the probability of each galaxy to host a \gls{cbc}. One simple approach models this as $p(\text{host}) \propto L_\text{K}^{\alpha}$, with $L_\text{K}$ the luminosity in the K-band. The work of \cite{Perna:2024lod} (see also \cite{Vijaykumar:2023bgs}) fits the parameter $\alpha$ appearing in the above power law to simulated data produced by \cite{Artale:2019tfl}. Their work finds $\alpha=9/4$ which would significantly down-weight faint galaxies.}

In the following, we assess whether the uncertainties of the reconstructed galaxy number counts are correct. 
To this end, we define a new random variable in the voxel $I$ as
\begin{equation}
\label{eq: def absolute deviation}
    \Deltapixel \deffrom \countGalaxiesTruepixel^{\rm True} - \langle\countGalaxiesTrueEstimatorpixel\rangle\,,
\end{equation}
and the relative deviations
\begin{equation}
\label{eq: def relative deviation}
    \Deltarelpixel \deffrom \frac{\Deltapixel}{{\rm std}(\countGalaxiesTrueEstimatorpixel)}\,,
\end{equation}
where $\langle\countGalaxiesTrueEstimator\rangle$ denotes the average over posterior samples in the number count $\countGalaxiesTrueEstimator$ and std is the standard deviation.
Finally, $\countGalaxiesTrue^{\rm True}$ is the true galaxy number count. 
Both of these random variables are computed for each voxel. 
The variable $\Deltapixel$ measures the absolute deviation between the true galaxy count (which is available since we use simulated data) and the mean of the reconstruction in that pixel. 
If our reconstructed number counts were systematically lower than their true values, $\Deltapixel$ would be positive and vice versa for an overestimation of the missing number counts.
While $\Deltapixel$ thus characterizes the mean number of reconstructed counts, it would not reflect any bias in the number count uncertainties. 
However, the second variable $\Deltarelpixel$ translates this in units of the standard deviation of the posterior. 
If the reconstruction correctly incorporates the model uncertainty (and \textit{if} the posterior in each voxel is close to a Gaussian distribution), $\Deltapixel$ should follow a normal distribution with zero mean and unit variance. Indeed, if most values of $\Deltarelpixel$ are close to zero, the standard deviation of the number counts is overestimated and the reconstructed catalog is conservative. If $\Deltarelpixel$ often reaches large values, where ``large'' depends on the number of voxels, this points to an overconfident reconstruction (i.e.~uncertainties too small). 
As such, $\Deltarelpixel$ is a diagnostic to validate the number count uncertainties on simulated data.

We plot the absolute deviations between the true and estimated galaxy number counts $\Deltapixel$ (as defined in Eq.~\eqref{eq: def absolute deviation}) from our reconstructed catalog in one $X$ slice in Fig.~\ref{fig: summary result phenom power spectrum} (bottom center panel);---few voxels show deviations above 60 counts and most voxels have values close to zero, as expected. 
If one divides these absolute deviations by their standard deviation, one obtains the lower-right panel, showing that few voxels have deviations above 8 standard deviations. 
If the posterior in each voxel was Gaussian, 8~sigma deviations should be extremely rare for $125,000$ voxels. Indeed, on average for Gaussian posteriors one should expect the maximum of the relative deviations at ${\rm inverf}(f)\sqrt{2} \approx 4.47$, with $f=1-1/50^3$ and $\rm inverf$ the inverse error function. However, in the present case, the posterior in each voxel is non-Gaussian and hence, one expects also larger relative deviations. 
Also note that the higher relative deviations are distributed homogeneously across the volume, indicating the independence of the reconstruction from the distance to the observer. 
Further validation and the comparison to the true number counts are carried out in the following section and Section~\ref{subsec: impact deep survey}.

\subsection{Impact of the assumed power spectrum} 
\label{subsec: comparison power spectrum}

To understand the importance of the assumed power spectrum that determines the \gls{dm} density contrast, we analyze the data with three different models for the \gls{dm} power spectrum: using ($i$) \cosmopowerjax{}\footnote{\change{For more details on this model, see the description in Sec.~\ref{subsec: large scale structure}. }} \cite{SpurioMancini:2021ppk} ($ii$) a phenomenological model and, ($iii$) a flat power spectrum. 
The purpose of varying the model of the DM power spectrum is to clarify which part of the data constrains the cosmological parameters. In principle, the cosmological parameters affect the luminosity distance-redshift relation and the \gls{dm} power spectrum. Whereas the luminosity distance-redshift relation is constrained via the observed redshifts and magnitudes, the \gls{dm} power spectrum is constrained more indirectly -- the observed galaxy number counts have to be related to the DM density contrast via the selection effect, and the bias parameters.  If this description is inadequate, this could potentially bias the inferred $H_0$ via the measured \gls{dm} power spectrum. 
Hence, the \cosmopowerjax{} model can constrain $H_0$ from the galaxy redshifts, magnitudes, and the power spectrum (since the power spectrum depends in this description on the cosmological parameters).
If this model recovers a biased $H_0$, one cannot determine whether the origin lies in a biased \gls{dm} power spectrum or a biased model of the magnitude distribution.
The phenomenological power spectrum model (model $ii$) does not depend on the cosmological parameters.
Therefore, if the inferred $H_0$ from the phenomenological power spectrum is biased, it has to result from the redshift and apparent magnitude data rather than a biased power spectrum (since the power spectrum is independent of the cosmological parameters for model $ii$).
Case ($iii$) assumes no correlation of the \gls{dm} density and hence, quantifies the impact of neglecting of spatial correlations between voxels for the reconstruction. 

\begin{figure}
    \centering
    \includegraphics[width=\textwidth]{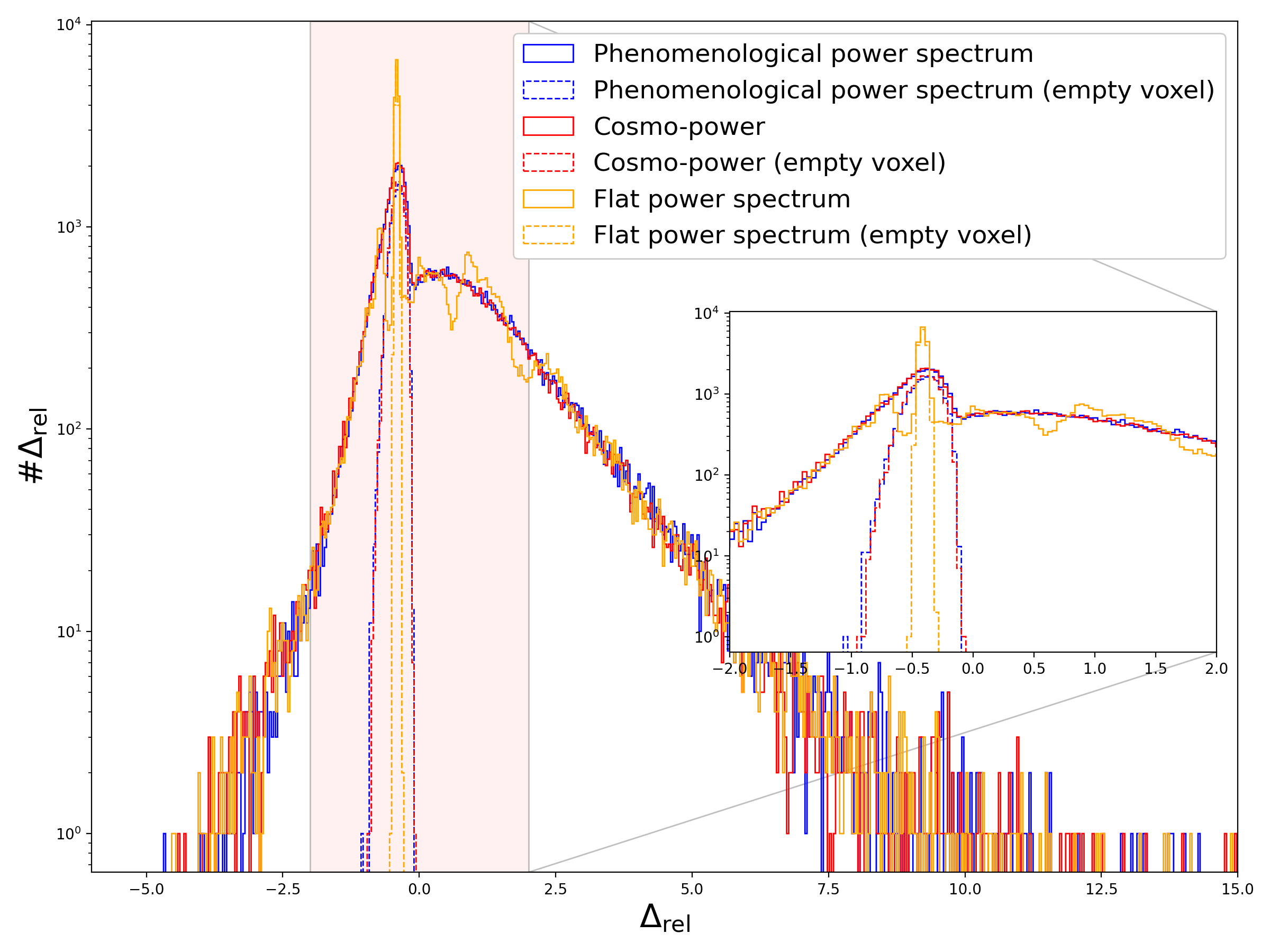}
    \caption{
    The histogram of the relative deviations, $\Deltarelpixel$, defined in Eq.~\eqref{eq: def relative deviation}, with a varying model for the power spectrum, as indicated in the legend. 
    For reference: for all $\Deltarelpixel > 0$, the true number counts are higher than the reconstructed number counts, i.e.~we underestimate the galaxy number counts. 
    Only few voxels show deviations that differ more than 8~standard deviations from the true values. 
    If the posterior in each voxel was Gaussian, this histogram would follow also a Gaussian distribution with mean zero and unit variance.
    The deviations do not show a strong dependence on the model that is used for the \gls{dm} density contrast, indicating that our analysis is data- rather than prior-driven. 
    In particular, whether we use a phenomenological power spectrum (blue) or a cosmology-informed power spectrum (red) does not matter. 
    Note the tail to large $\Deltarelpixel$ values, meaning that our model underestimates the galaxy number counts.
    We also draw the deviations conditioned on that voxel containing no galaxies (dashed lines; ``empty voxel''). 
    This sub-population dominates the overall deviations near zero, and is biased towards $\Deltarelpixel<0$, since it is impossible to underestimate a vanishing number count. 
    }
    \label{fig: relative deviation comparison}
\end{figure}

We focus on a galaxy-cube with side length 150~Mpc$/h$, divided in 40 segments per dimension and impose an apparent magnitude threshold of 21 on all detected galaxies.
From this simulated data, we run the inference outlined above and obtain posterior samples in $\hyperparams$, $\densityDMpixel$ and the estimated galaxy number count $\countGalaxiesTruepixel$. 
To verify the reliability of the reconstruction, we make use of the true galaxy number counts and compute the relative deviations as defined in Eq.~\eqref{eq: def relative deviation}. 
We then histogram the relative deviations $\Deltarelpixel$ for the $40^3$ voxels in Fig.~\ref{fig: relative deviation comparison}. 
If the individual posterior distribution of one voxel, $\p(\countGalaxiesTruepixel|\Data)$, were Gaussian, the histogram should closely follow a Gaussian distribution with unit variance and mean zero.
If our posterior uncertainty was too small, the variance of $\Deltarelpixel$ should be larger than one, and vice versa for underestimated uncertainties.
If one underestimates the galaxy number counts, the mean of $\Deltarelpixel$ shifts to the right and similarly, an overestimate leads to a left-shift. 
Unfortunately, for non-Gaussian posterior distributions, it quickly becomes intractable to obtain closed-form expressions for this histogram and our situation is further complicated since the single-pixel posteriors are correlated.
As such, Fig.~\ref{fig: relative deviation comparison} provides a simplified diagnostic to assess the reliability of our reconstruction and its uncertainty.

In Fig.~\ref{fig: relative deviation comparison} we compare the histogram of the relative deviations with three different models for the power spectra introduced above: a cosmologically-informed power spectrum (model $i$), a phenomenological power spectrum (model $ii$), and a flat power spectrum (model $iii$). 
The overall structure of all three histograms agrees; all peak at relative deviations close to zero, indicating that most voxels are reconstructed with small differences from the underlying truth. 
This indicates that the assumed power spectrum has little influence on the shape of the relative deviations, given that the phenomenological and the cosmologically-motivated power spectrum give virtually identical results. 
Therefore, we conclude that the results are data-driven and are not strongly impacted by the model assumptions on the underlying correlation structure for the \gls{dm} density contrast.
However, using a flat power spectrum (assuming no large-scale structure correlations) leads to a deviation from the other two assumptions for deviations close to zero. 
Also note that both non-flat power spectrum models show peaks close to zero. One can condition the relative deviations to only include empty voxels, represented as dashed curves. From this subset of deviations it becomes apparent why these voxels only contribute to negative values: one cannot underestimate empty voxels (cf.~Eq.~\eqref{eq: def relative deviation} for $\countGalaxiesTruepixel^{\rm True} = 0$) and hence, $\Deltarelpixel\leq 0$.

Let us further elaborate on the asymmetry of this histogram, since our posterior tends to reconstruct the number counts at lower number counts than their true values ($\Deltarelpixel > 0$). 
We consider a simpler toy model for $\Deltarelpixel$, where the observations are one single number count $n$ that is Poisson distributed for a given rate $\mu_{\rm true}$, i.e.~$\p(n|\mu_{\rm true})=\Poisson(\lambda=\mu_{\rm true}; k=n)$. 
In that simplified setup (fixing the true rate $\mu_{\rm true}$ and assuming a flat prior on $n$), and given an observation of $n$, the mean and standard deviation of the posterior are both $\langle \mu \rangle = {\rm std}(\mu ) = n + 1$.  
Then, one can compute the relative deviations via Eq.~\eqref{eq: def relative deviation}) for the estimated rate which gives $\Deltarelpixel=(\mu_{\rm true} - n+1) / \sqrt{n+1}$. 
From this, one can calculate numerically the probability distribution of $\p(\Deltarelpixel|\mu_{\rm true})$ (marginalizing over all possible observations $n\in\mathbb{N}_{\geq 0}$) which gives same lopsidedness as we have found in the more realistic case. 
This lopsidedness of the $\Deltarelpixel$ histogram is thus a direct result from the Poisson-nature of the problem, rather than indicating a systematic bias in the reconstruction.

\subsection{Impact of redshift uncertainties}
\label{subsec: impact redshift uncertainties}

\begin{figure}
    \centering
    \includegraphics[width=\linewidth]{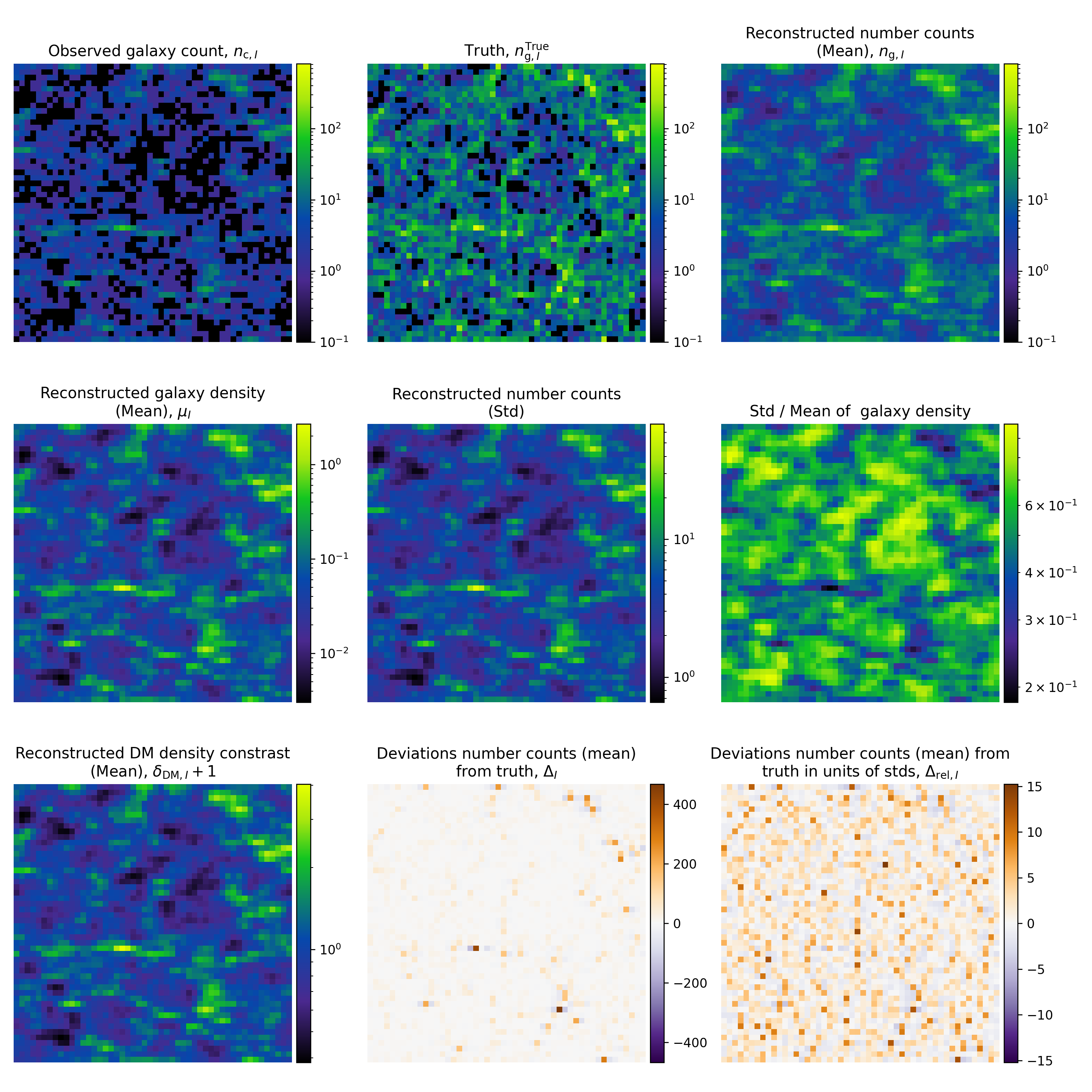}
    \caption{
    A slice perpendicular to the $X$-axis summarizing the reconstructed Millennium catalog using a phenomenological power spectrum for the \gls{dm} density contrast. The observed counts were obtained via binning of the galaxies according to their measured redshifts that include redshift uncertainties (cf.~Eq.~\eqref{eq: redshift error}). 
    Since the observer is located at the left the horizontal axis corresponds to the (measured) redshift coordinate (under the flat-sky approximation). 
    For the description of the individual panels, see Fig.~\ref{fig: summary result phenom power spectrum}. 
    Since we neglect redshift uncertainties during inference, the galaxy density and the \gls{dm} density both are reconstructed incorrectly and include the spurious structures along the line-of-sight (which are not present in the true densities). These structures also propagate to the deviations from the true number counts: at strong overdensities, we have $\Deltarelpixel>0$, indicating that we underestimate the number count, whereas the neighboring voxels to the right and left are negative, indicating an overestimate. 
    This overestimate-underestimate-overestimate (blue-red-blue) is a direct effect of the spilling of galaxies in left and right voxels through their measured redshift. 
    }
    \label{fig: summary redshift errors fine grid}
\end{figure}

Above, we have assumed redshift measurements with no uncertainty. The following section discusses the effects of this assumption on the overall reconstruction.
At the present voxel length of $5$~Mpc$/h$, we find while the analysis neglecting redshift uncertainties succeeds in reconstructing the smoothed out galaxy number count correctly, the redshift uncertainties lead to a broader distribution of the relative deviations $\Deltarelpixel$, i.e.~lead to an underestimation of the number count uncertainties. Let us now detail this result. 

Throughout this subsection, we assume for the data generation that the measured redshifts follow a Gaussian distribution centered at the true redshift, $z$, with a standard deviation of \cite{EUCLID:2011zbd}
\begin{equation}
\label{eq: redshift error}
    \sigma(z) = 0.001\times(1 + z)\,.
\end{equation}
This model is compatible with forecasts for the spectroscopic redshift uncertainties from Euclid \cite{EUCLID:2011zbd, Euclid:2021icp}. 
The redshift uncertainties lead to a ``spilling'' of galaxies from one bin to its neighbors, smoothing out the observed large-scale structure along the line-of-sight, as shown in Fig.~\ref{fig: summary redshift errors fine grid}.
With the redshift uncertainties causing this spilling between redshift bins, we build a magnitude-limited survey ($\mthresh=17$) for a galaxy cube of side length 250~Mpc$/h$ and 50 segments per dimension.
Note that the observed apparent magnitudes are computed via their absolute magnitudes and their true redshift, i.e.~the apparent magnitudes are not uncertain.

From this simulated survey, we infer the number counts with the method specified in Eq.~\eqref{eq: likelihood data with ng depedency} and Eq.~\eqref{eq: full likelihood marginalized over unobserved galaxy number counts}, i.e.~we neglect redshift uncertainties in our modeling. For the description of the DM power spectrum, we use the phenomenological model defined in Eq.~\eqref{eq: def phenomenological description power spectrum}. 
The reconstruction is then validated against the true galaxy number counts binned according to their true redshift. 
Fig.~\ref{fig: summary redshift errors fine grid} illustrates the result of the reconstruction;---while the estimated galaxy number counts (top-right panel) correspond closely to the true galaxy number count (top-center panel), they retain the spurious correlations along the line-of-sight. 
Especially near strong galaxy overdensities, the relative deviations, $\Deltarelpixel$ (bottom-right panel, cf.~Eq~\eqref{eq: def relative deviation}), reach values above 15, indicating that the inferred number counts are many standard deviations away from their true values. 
We thus identify the limitation of our current approach. If the voxel size is comparable to or below the redshift uncertainty (i.e.~$\Delta z_{\rm pixel}\leq\mathcal{O}(2)\Delta z$, with $\Delta z_{\rm voxel}$ the voxel length in redshift and $\Delta z$ the redshift uncertainty), the reconstruction starts to degrade and the number count uncertainties are underestimated.

Neglecting the redshift uncertainties also leads to a strong bias of the inferred slope of the \gls{dm} power spectrum $\PSn$, since structure at small scales is smoothed out. When analyzing the data with the model neglecting redshift uncertainties, we find $\PSn=-1.438^{+0.002}_{-0.003}$, with the lower and upper 1~sigma intervals indicated. The inferred power spectrum peaks thus at lower $k$ than the power spectrum obtained when analyzing data without redshift uncertainties. 
We also find that $H_0$ is inferred at small values, further pointing to the necessity of including redshift uncertainties in our model. 

The likelihood that marginalizes over the galaxy number counts (cf.~Eq.~\eqref{eq: full likelihood marginalized over unobserved galaxy number counts}) can be naturally extended to incorporate redshift uncertainties without introducing a significantly larger computational burden.
In this case, the rate is modified according to $\mu(z,\skypos,M|\hyperparams,\densityDM) \rightarrow \mu(\hat z,\skypos,M|\hyperparams,\densityDM)$, with $\hat z$ the \textit{measured} redshift, $\hat z$, (including redshift uncertainties), that is given by 
\begin{equation}
    \mu(\hat z,\skypos,M|\hyperparams, \densityDM) = 
    \int \dd z~\mu(z,\skypos,M|\hyperparams, \densityDM)\,\p(\hat z|z) \,,
\end{equation}
where we approximate $\p(\hat z|z) \approx \mathcal{N}(\mu=z,\sigma = 0.001;\hat z)$. 
The above expression is effectively a convolution of the rate in true redshift with a Gaussian kernel that smooths out the density fluctuations along the line-of-sight. 
Analyzing the simulated catalog with this likelihood, we obtain a slope of $\PSn=-0.63^{+0.02}_{-0.02}$ for the power law exponent of the power spectrum. This inferred $\PSn$ value is much closer to $\PSn$ from the analysis with data without redshift uncertainties. Additionally, the bias of the inferred $H_0$ also disappears with this approach.   
However, to obtain posterior samples in the galaxy number counts, one has to formulate the likelihood that does not marginalize over the galaxy number counts, and that accounts for redshift uncertainties (the analog of Eq.~\eqref{eq: likelihood data with ng depedency}). This is a complex expression that requires the modeling of number count spilling between different voxels, the implementation of which we leave to future work.

\subsection{Improved magnitude distribution from a smaller, deeper galaxy survey}
\label{subsec: impact deep survey}

\begin{figure}[htbp]
    \centering
    \includegraphics[width=\textwidth]{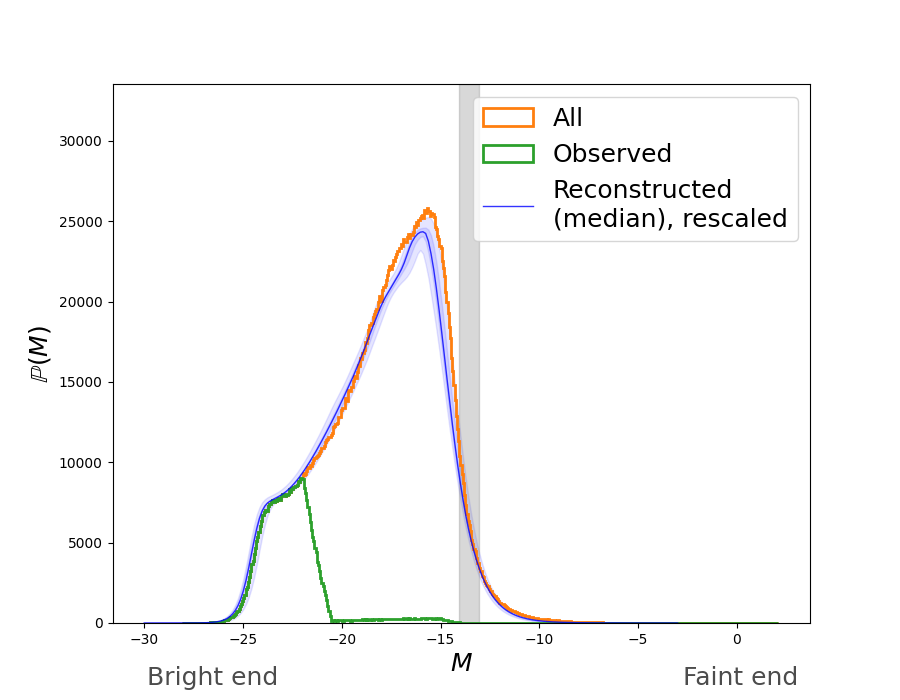}
    \label{fig:with_deep_survey}
    \caption{The reconstructed magnitude distributions with a deeper galaxy survey for a portion of the sky, to be compared to the reconstructed magnitude distribution without the deep survey, cf.~Fig.~\ref{fig: magnitudes reconstructed}. 
    The median of the reconstructed distribution is drawn as a blue line, with the 1~sigma and 2~sigma intervals indicated as shaded regions. 
    We rescale the inferred magnitude distribution such that it coincides with the observed magnitude distribution at the bright end to provide an easier visual comparison. 
    We have marked the faintest galaxy observed with a grey band. 
    The 25 additional sky pixels (with an apparent magnitude threshold of 24 instead of 17) that provide much fainter galaxies strongly impact the measurement of the magnitude distribution at the faint end. 
    }
    \label{fig: comparison deeper survey}
\end{figure}

\begin{figure}
    \centering
    \includegraphics[width=\textwidth]{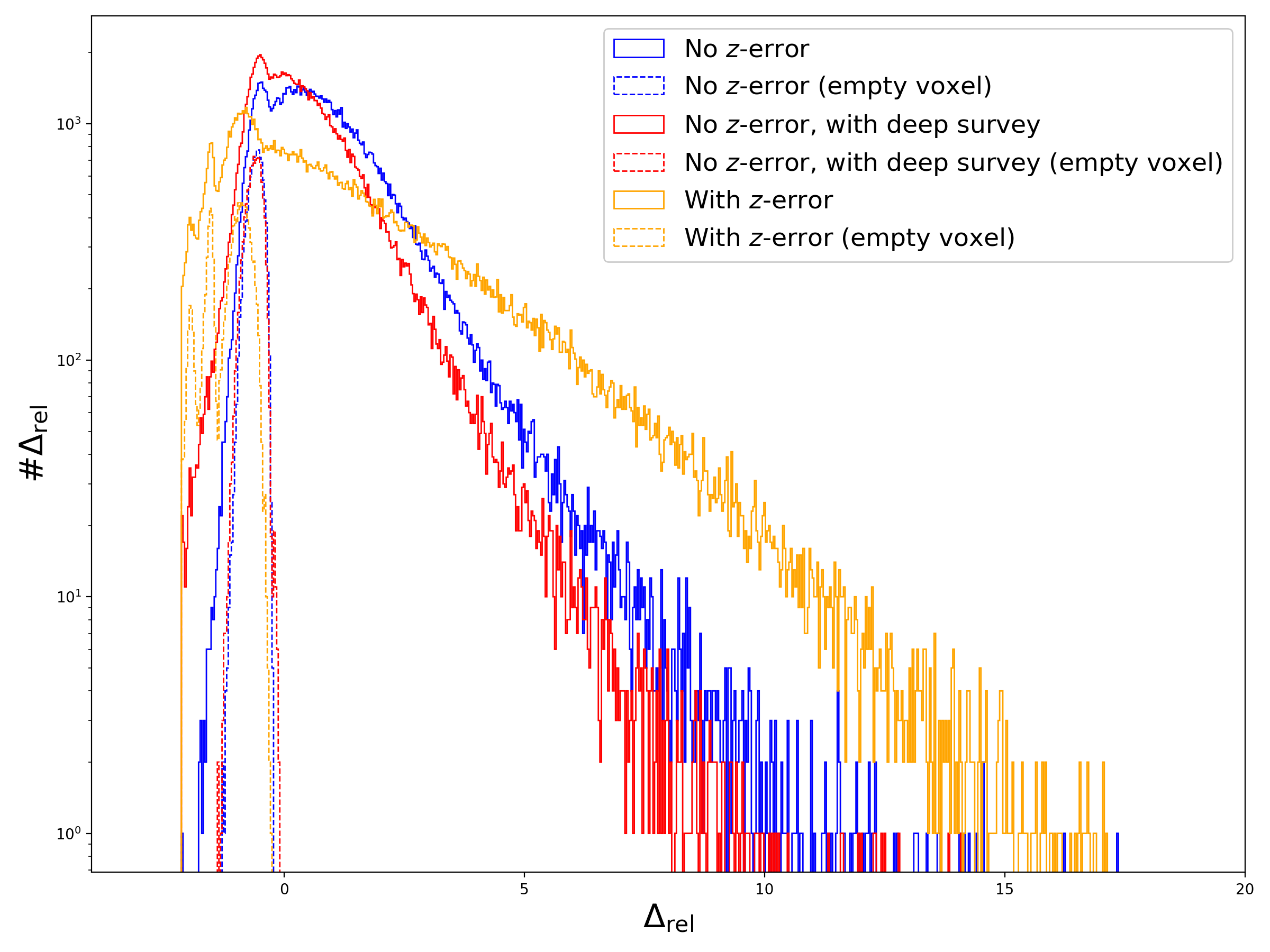}
    \caption{
    The histogram of the relative deviation variable $\Deltarelpixel$, defined in Eq.~\eqref{eq: def relative deviation} with (red) and without (blue) an improved apparent magnitude threshold from 17 to 24 for 25 sky pixels. The yellow histogram shows the relative deviations for an analysis with data including redshift uncertainties (cf.~Sec.~\ref{subsec: impact redshift uncertainties}). 
    For reference: for all $\Deltarelpixel > 0$, the true number counts are higher than the reconstructed number counts, i.e.~we underestimate the galaxy number counts. 
    The reconstructions analyzing data without redshift uncertainties show the same behavior as Fig.~\ref{fig: relative deviation comparison} and have only a few outliers above 10~standard deviations.
    The result using the partially deeper survey (red) has fewer values at large deviations, indicating that the observation of faint magnitudes aids a more faithful reconstruction. 
    The analysis with redshift uncertainties (but neglecting these during the analysis) leads to a broader $\Deltarelpixel$-histogram that implies that number count uncertainties are underestimated. The relative deviations in this case for empty voxels (yellow dashed line) are biased towards negative values;---spilling from neighbouring voxels caused by redshift uncertainties leads to an overestimate of the reconstructed number counts ($\Deltarelpixel<0$, cf.~Eq.~\eqref{eq: def relative deviation}).   
    }
    \label{fig: relative deviation comparison deep survey}
\end{figure}

In the previous sections, we discussed that the inferred magnitude distribution for magnitudes fainter than the faintest observed galaxies can be incompatible with the true $\p(M|\hypermagnitudes)$. To more reliably reconstruct the faint part of the magnitude distribution, we require data that provides information on these faint galaxies. We simulate a galaxy survey that is significantly deeper in a small portion of the sky -- one can imagine this as a rough analogy of, say, the Euclid Deep Survey ($\sim 53$ deg$^2$, compared to $15,000$ deg$^2$ for the Euclid Wide Survey \cite{Euclid:2021icp}). 
We restrict the Millennium Simulation to a comoving distance box of $250$~Mpc$/h$ in size, with a pixelization of 50 per dimension. As the baseline survey, we consider $2475=2500-25$ sky pixels to have an apparent magnitude threshold of 17. The remaining $25$ pixels ($1\%$ of the sky) have an improved apparent magnitude threshold of 24. To provide a fair comparison, we leave all other parameters and their prior ranges the same, using a phenomenological model for the power spectrum.  
Effectively, this setup provides a much more complete survey in a narrow square rectangular box. Although not many more galaxies are observed, from 450563 to 476396, the magnitude of the faintest galaxy included in the survey is shifted from -20.5 to -13.6, a significant improvement that constraints the faint end of the magnitude distribution.  
 
Fig.~\ref{fig: comparison deeper survey} shows the reconstruction of the magnitude distribution with the deep survey and should be compared to Fig.~\ref{fig: magnitudes reconstructed}. 
The baseline magnitude distribution recovers the true distribution faithfully but differs for faint magnitudes as anticipated. In contrast to this, the reconstruction using the deep survey only has very few magnitudes above $M\geq -20$, but these provide sufficient information to recover almost the entirety of the magnitude distribution.  The increased depth to faint magnitudes emphasizes the importance of deep surveys, allowing for an improved number count reconstruction.  Note that we assume the $1\%$ sky pixels to be representative of the entire magnitude distribution. If this is not the case, the reconstructed magnitude distribution will generally be biased.  As discussed in Sec.~\ref{subsec: results vanilla case}, one important parameter of the magnitude model that impacts the overall reconstruction is the fraction of unseen galaxies, $\muFraction$, at a fixed absolute magnitude $\muThresholdAbove$. As a reminder, we construct the magnitude distribution such that all magnitudes above $\muThresholdAbove$ take up the relative fraction $\muFraction$ of the whole distribution. As such, $\muFraction$ does not impact the shape of the distribution but rather how many galaxies are faint. If $\muFraction$ is fixed to an incorrect value, the magnitude distribution in the regime of bright galaxies, where we have observations, is recovered without bias. However, if $\muFraction$ is too large, the relative number of faint galaxies is overestimated and vice-versa for a low $\muFraction$. This biased recovery translates into an incorrect overall estimate of the number counts. 
To more quantitatively compare the two reconstructions, Fig.~\ref{fig: relative deviation comparison deep survey} shows the relative deviations for the two cases. Since the baseline survey estimates $\muFraction$ roughly correctly, the overall reconstructions with and without the deep survey are similar, but not identical. 
The deeper survey through its improved reconstruction of the faint magnitude distribution leads to fewer deviations $\Deltarelpixel$ and hence, indicates an also improved reconstruction of the galaxy number counts. 
As previously seen in Sec.~\ref{subsec: comparison power spectrum}, we find the tendency of $\Deltarelpixel > 0$ (cf.~Eq.~\eqref{eq: def relative deviation}).

To conclude, even small patches of the sky with a higher apparent magnitude threshold allow for a significantly improved reconstruction of the magnitude distribution. This is achieved since the faintest galaxy included in the survey (in absolute magnitude) is significantly increased towards larger magnitudes and hence provides information in that previously unconstrained region of magnitude space. If the fraction of faint galaxies is recovered correctly, the reconstructions with and without the deeper survey agree reasonably well.

\section{Conclusions}

In this study we presented a novel approach for reconstructing galaxy catalogs, which is designed to improve the measurement of cosmological parameters, particularly the Hubble constant $H_0$, with gravitational waves. 
Our method leverages the large-scale structure correlations to overcome the limitations posed by incomplete galaxy catalogs. 
Using a Bayesian framework, we performed robust reconstructions of galaxy number counts on simulated data. This framework allows for the simultaneous estimation of cosmological parameters, the parameters describing the distribution of absolute magnitudes of galaxies, the parameters of the dark matter power spectrum and the bias parameters that link the DM density to the galaxy density.

Our approach has several associated challenges. 
($i$) Since we infer the galaxy density and \gls{dm} density contrast in each voxel, the problem is described by a high-dimensional posterior. ($ii$) The modeling has a preferred ``direction'': it is straightforward to compute an observed galaxy catalog from a \gls{dm} density contrast, but difficult to solve for the inverse. 
($iii$) The catalog's magnitude limit causes many \gls{dm} density contrast realizations mapping to the same observation, leading to degeneracies in the posterior. To address the problem of high-dimensionality we employed the autodifferentiable programming language JAX \cite{jax2018github} and a Hamiltonian Monte-Carlo sampling algorithm, relying on probabilistic programming using \numpyro{} \cite{Phan:2019elc}. 
This allowed us to in produce posterior samples that can quantify the uncertainty of the reconstructed galaxy counts in each voxel.
The remaining challenges were tackled with a forward-modeling philosophy -- for given cosmological parameters, the phenomenological parameters describing the matter power spectrum, the magnitude distribution parameters and the galaxy bias parameters -- one produces a possible realization of the galaxy field. This galaxy field realization then allows to compute the probability of having measured the magnitude-limited survey at hand.

We validated our methodology using the Millennium Simulation, demonstrating its accuracy in reconstructing the galaxy field. 
We found that while the bright end of the magnitude distribution is accurately recovered, the faint end shows larger discrepancies, indicating the limitations of our approach: the reconstructed magnitude distribution cannot be trusted above the faintest (absolute) magnitude that is observed. This limitation clearly points to the necessity of having a deep survey to obtain faint magnitudes;---we have demonstrated that even a deeper survey limited to $1\%$ of all sky pixels significantly aids the reconstruction of the magnitude distribution. 
Furthermore, our analysis revealed that the choice of the power spectrum model for the \gls{dm} density contrast does not significantly impact the reconstruction results, highlighting the robustness of our data-driven approach at the scales considered here. However, in certain cases, the model tends to overestimate galaxy counts in high-density regions, suggesting potential future improvement for the description of the \gls{dm} density contrast. 

Additionally, we discussed the impact of redshift uncertainties on the reconstruction process. If the redshift uncertainties move galaxies over distances comparable to the voxel resolution these uncertainties cannot be neglected and will lead to an underestimate of the number count uncertainties. This indicates our methods can be applied immediately to spectroscopic galaxy surveys with the voxel length larger than the redshift error, but will require adaptation for photometric surveys, where typical redshift errors are larger; we leave this to future investigation.

There are several improvements needed for future work. 
For the reconstruction of real surveys, the flat-sky approximation should be replaced by a more realistic distance measure and the \gls{dm} density contrast prior should include effects resulting from the survey being observed on the light-cone. 
Despite this, our method shows a strong potential for application in gravitational wave cosmology, where redshift information of the \gls{gw} host is crucial for a precise $H_0$ measurement.

In summary, this Bayesian approach represents a significant advancement over previous methods, providing a more complete and reliable reconstruction of galaxy catalogs. 
Our reconstructions will allow us to extend the informative reach of the line-of-sight priors that are essential to the galaxy catalog (a.k.a. dark sirens) method.
In this way we have laid the groundwork for improved cosmological parameter estimation, with promising implications for cosmological inference with \gls{gw} observations and galaxy catalogs. Future work will focus on refining the magnitude distribution model, incorporating redshift uncertainties, allowing this methodology to measure cosmological parameters from real data.

\section*{Acknowledgments}
We thank Sesh Nadathur, Coleman Krawczyk, Bartolomeo Fiorini, Surhud More, Rachel Gray, Simone Mastrogiovanni, Rudy Morel and Charles Dalang for helpful discussion. K.L. and T.B. are supported by ERC Starting Grant SHADE (grant no.\,StG 949572). T. B. is further supported by a Royal Society University Research Fellowship (grant no.\,URF$\backslash$R$\backslash$231006).
Numerical computations were carried out on the \texttt{Sciama} High Performance Computing (HPC) cluster, which is supported by the Institute of Cosmology and Gravitation (ICG), the South-East Physics Network (SEPNet) and the University of Portsmouth.

\appendix

\section{Generation of log-normal fields with a target power spectrum}
\label{app: generation log-normal field steps}

In Sec.~\ref{subsec: matching the 2-point statistics}, we describe the matching of the power spectra of a log-normal field and its underlying Gaussian random field. 
We rely on the following steps to use a Gaussian random field realization to generate a log-normal field with a target power spectrum. 

\begin{enumerate}
    \item Fourier-transform the target power spectrum to obtain the spatial correlation function. 
    \item Use Eq.~\eqref{eq: relation spatial correlation function log-normal and gaussian random field} to link the spatial correlation function of the log-normal field with the spatial correlation function of the underlying \gls{grf}. 
    \item Fourier-transform the spatial correlation function $\spatialcorrelationfieldgaussian$ to obtain the 2-point statistics of the \gls{grf}, denoted as $P_{\rm eff}^{(K)}$. 
    \item Draw a field realization of a white noise \gls{grf} in spatial domain $\fieldgausswhitenedspatial$ (flat power spectrum, with zero mean and unit variance). 
    \item Fourier-transform this field to obtain the white noise Gaussian field $\fieldgausswhitenedfourier$ in Fourier space. 
    \item Reweight the field in Fourier space according to $\fieldgausscoloredfourier\deffrom \sqrt{P_{\rm eff}^{(K)}}\,\fieldgausswhitenedfourier$. 
    \item Fourier-transform $\fieldgausscoloredfourier$ to obtain $\fieldgausscoloredspatial$
    \item Apply the transformation of Eq.~\eqref{eq: def log-normal field} to obtain $\densityDM$. 
\end{enumerate}
One can avoid point 4 by directly drawing Gaussian field realizations in Fourier space. However, the field is real and therefore respects the symmetry $\fieldgausswhitenedfourier(k) = \fieldgausswhitenedfourier(-k)^* $ in Fourier space that has to be included during sampling.

\section{Definition of distributions}
\label{app: definitions distributions}

We follow the standard definitions of the following distributions, given by
\begin{itemize}
    \item Normal distribution
    \begin{equation}
       \mathcal{N}(\mu, \sigma; x) = \frac{1}{\sqrt{2\pi\sigma^2}} \exp\left(-\frac{(x - \mu)^2}{2\sigma^2}\right)\,,
\end{equation}
\item Poisson Distribution
\begin{equation}
\Poisson(\lambda; k) = \frac{e^{-\lambda} \lambda^k}{k!} \,,
\end{equation}
\item Binomial Distribution
\begin{equation}
\Binomial(n, p; k) = \binom{n}{k} p^k (1-p)^{n-k} \,.
\end{equation}
\end{itemize}

\section{Magnitude distribution}
\label{app: magnitude distribution}

We model the magnitude distribution $\p(M|\hypermagnitudes)$ as a succession of transformations of a random variable $\chi$, that follows (by definition) a uniform distribution between $0$ and $1$, i.e.
\begin{equation}
    M \deffrom t_{N} \circ t_{N-1} \circ \ldots \circ t_2 \circ t_1 (\chi) \,,
\end{equation}
where $t_i$ is the $i$th sub-transformation and the number of sub-transformations is $N$. 
As building blocks for these sub-transformations, we need the following transformations. The sigmoid transformation, defined as
\begin{equation}
    \sigma(x) \deffrom \frac{1}{1 + \exp(-x)}\,,
\end{equation}
and the softplus, given by
\begin{equation}
    {\rm Softplus}(x) \deffrom {\rm log}\left[1 + \exp (x)\right]\,.
\end{equation}
We have also constructed a third-order polynomial that is injective and monotonically increasing on the interval $\left[0,1\right]$, with the function and its inverse defined as
\begin{align}
    \mathcal{P}_3(x; a_i, b_i) 
    &\deffrom
    \frac{x \left( 3(a_i^2 + b_i^2) - 3a_i x + x^2 \right)}{1 - 3a_i + 3(a_i^2 + b_i^2)}\,,
    \\
    \nonumber
    \mathcal{P}_3^{-1}(x; a_i, b_i)
    &\deffrom
    a_i - \frac{{3  2^{\frac{1}{3}}  b_i^2}}{{(f(x) + r(x))^{\frac{1}{3}}}} + \frac{{(f(x) + r(x))^{\frac{1}{3}}}}{{3  2^{\frac{1}{3}}}}
    \,, \\
    \nonumber
    f(x) &\deffrom 
     -27  a_i^3 - 81  a_i  b_i^2 + 27  x - 81  a_i  x + 81  a_i^2  x + 81  b_i^2  x \,,
    \\ 
    \nonumber
    r(x)&\deffrom 
    \sqrt{2916  b_i^6 + f^2}\,,
\end{align}
where $r$ stands for root. 
We set the transformations as follows
\begin{align}
    t_i &= \mathcal{P}_3^{-1}(x;a_i, b_i) \quad \text{for $i\in\{0,2,4\}$ } \,,
    \\
    \nonumber
    t_i &= \mathcal{P}_3(x;a_i, b_i) \quad \text{for $i\in\{1,3,5\}$ } \,,
    \\
    \nonumber
    t_6 &= 
    \sigma^{-1}(x)
    \,,
    \\
    \nonumber
    t_7 &= \epsilon x + \epsilon_2
    \\
    \nonumber
    t_8 &= 
    {\rm Softplus}(x)
    \,,
    \\
    \nonumber
    t_9 &=
    \sigma x + \mu\,.
\end{align}
Finally, we define a magnitude value $\muThresholdAbove$ above which only a fraction $\muFraction$ of galaxies is observed. This accounts for the fact that the faint end of the galaxies is not observed, and hence we cannot place any meaningful constraints on the parameters governing the faint end of the galaxy magnitude distribution, i.e.
\begin{equation}
    \int_{-\infty}^{\muThresholdAbove}
    \p(M|\hypermagnitudes)\dd M \overset{ }{=} 1 - \muFraction\,.
\end{equation}
We ensure the above constraint by the sub-transformation
\begin{equation}
    t_{10} = x + (\muThresholdAbove - \mu_{\rm eff})\,,
\end{equation}
with $\mu_{\rm eff} \deffrom (t_9 \circ t_8 \circ  \ldots\circ t_0)(1-\muFraction)$. 
Note that the parameters $\muThresholdAbove$ and $\muFraction$ are degenerate and we only sample in the parameter $\muFraction$. 
As a final step, we apply a transformation $t_{11}$ that ensures that the magnitude parameters as defined above do not correlate with $H_0$, given by 
\begin{equation}
    t_{11}(x) \deffrom x + 
    5 \log_{10}\left(
    \frac{H_0}{H_{0,{\rm ref}}}
    \right) \,,
\end{equation}
for $H_{0,{\rm ref}}$ a reference value of the Hubble constant that is fixed prior to the analysis. To see whether this expression shifts the distribution appropriately, let us consider the following a galaxy that is observed with a redshift and an apparent magnitude. 
Increasing $H_0$ leads to a smaller inferred luminosity distance, consequently to a \textit{higher} associated absolute magnitude (one infers that the galaxy must be fainter). Hence, the absolute magnitude distribution should be shifted to the right, exactly as the above expression indicates.

\begin{figure}
    \centering
    \includegraphics[width=\textwidth]{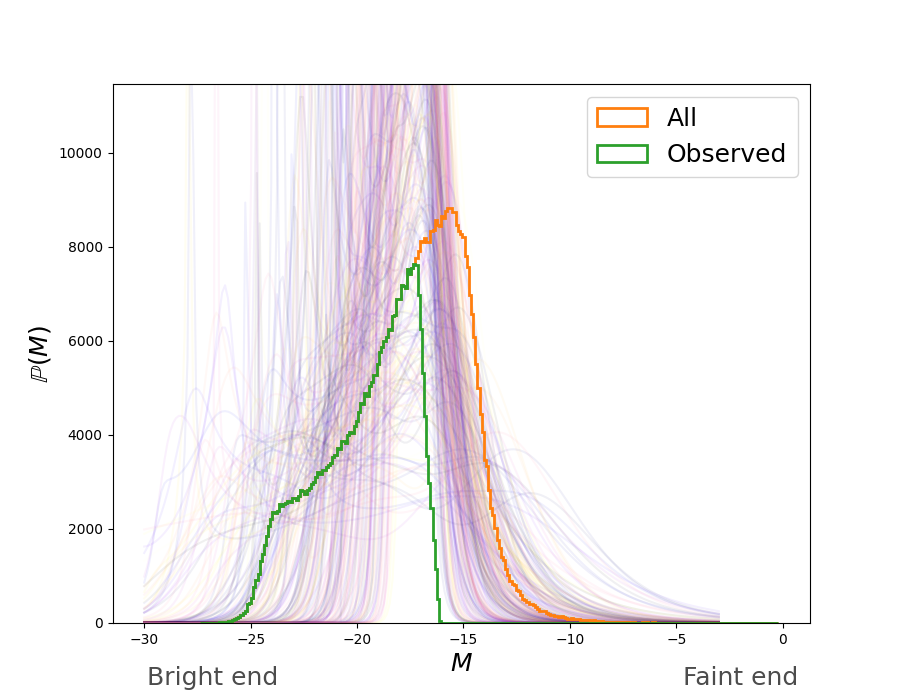}
    \caption{The magnitude distribution of all and detected galaxies. We also plot the magnitude distribution model as a set of semi-transparent lines to show the wide variety of different distributions that one can fit with the model. The color of the curves corresponds to the Hubble constant. All parameters, $\hypermagnitudes$ and $H_0$ are drawn from the prior. }
    \label{fig: example magnitude model}
\end{figure}

All in all, we thus have 18=12+2+2+1+1 parameters describing the magnitude distribution. 
This particular form was chosen to match the power law increase for low magnitudes, followed by an exponential decay for faint ones. (Compare to the distribution of magnitudes of Fig.~\ref{fig: magnitudes reconstructed}). 
To show the versatility of our transformations, we draw $\hypermagnitudes$ from the prior we consider in this work and plot the resulting magnitude distributions in Fig.~\ref{fig: example magnitude model}.
We have also verified that the results do not strongly depend on the number of transformations used. 

\section{Derivation of the hierarchical likelihood}
\label{app: derivation hierarchical likelihood}

In the following, we derive the hierarchical likelihood for observing $n_C$ data points (in a first step without imposing a selection bias on the data). 
For simplicity, we assume a finite discretization of the entire data space, with $N$ elements spanning all possible data and we denote the (finite) volume element of the data as $\finiteData{}$. 
We also assume that the rate $\rateAll{}$ is constant over the domain of one data bin $\Data_i$.
One can obtain the infinitesimal version of the main result (cf.~Eq.~\eqref{eq: hierarchical likelihood data finite vol}) by considering the limit $N\rightarrow \infty$, $\finiteData{}\rightarrow0$, but $N\finiteData{}=\text{constant}$, where the constant is independent of the exact choice of $N$.

We begin by focusing on one of the $N$ bins (associated to the data bin $\Data$), where we assume that the number of sources is Poisson distributed, i.e.
\begin{equation}
    \p(n\text{ observations with data }\Data|\hyperparams, \densityDM) 
    =
    e^{-\rateAll{}\,\finiteData{}}\;\frac{(\rateAll{}\finiteData{})^n}{n!}\,.
\end{equation}
This expression can then be used to compute the likelihood of observing number counts in \textit{all} $N$ different bins
\begin{align}
\label{eq: general hierarchical likelihood 2}
    \p(\{n_i\}_{i=1,2,\ldots,N}|\hyperparams,\densityDM) &
    = 
    \prod_{i=1}^N 
    \p(n_i\text{ observations with data }\Data_i|\hyperparams, \densityDM)
    \\
    \nonumber
    &
    =
    e^{-\sum_{i=1}^N\mu(\Data_i|\hyperparams,\densityDM)\,\finiteData{i}}\;\prod_{i=1}^N
    \Biggl\{
    \frac{\left[\rateAll{i}\,\finiteData{i}\right]^{n_i}}{n_i!}
    \Biggr\}
    \,,
\end{align}
If all observed data are distinct, one can pick the discretization that is sufficiently fine, so that $n_i=1$ for a selected subset of all data bin indices, denoted as $C$, and the rest of $n_i=0$ for $i\notin C$. This implies that $\sum_{i=1}^Nn_i=n_C$, for $n_C$ the number of elements in $C$.
\begin{equation}
\label{eq: hierarchical likelihood counts}
    \p(\{n_i\}_{i=1,2,\ldots,N}|\hyperparams, \densityDM)
    =
    e^{-\sum_{i=1}^N\mu(\Data_i|\hyperparams,\densityDM)\,\finiteData{i}}\prod_{i=1}^{n_C}
    \rateAll{i}\,\finiteData{i}
    \,.
\end{equation}
Usually, we want to compute the likelihood normalized with respect to the data space. We thus consider the variable change that replaces the list of event counts in each data bin, $\{n_i\}_{i=1,2,\ldots,N}$, with a list of possible data, that we denote as $\{\Data_i\}_{c\in C}$. 
The two likelihoods are then related by
\begin{equation}
    \p(\{n_i\}_{i=1,2,\ldots,N}|\hyperparams, \densityDM)
    =
    \p(\{\Data\}_{c\in C}|\hyperparams, \densityDM)
    \left[
    \prod_{i=1}^{n_C}\finiteData{i}
    \right]\,
    \overcount
    \,,
\end{equation}
where the factor $\overcount$ accounts for overcounting; if one exchanges two data $\Data_{i}\leftrightarrow \Data_j$, the number count is invariant. 
Hence, there is a many-to-one mapping from possible data-lists $\{\Data\}_{c\in C}$ to the number counts. Indeed, when integrating over all possible data, there are 
\begin{equation}
    \overcount = \frac{\left(\sum_{i=1}^N n_i\right)!}{\prod_{i=1}^N (n_i!)} = \left(\sum_{i=1}^N n_i\right)! = n_C! \,,
\end{equation}
possible data configurations that lead to the same number count configuration.
The second equality holds, since we have assumed that $n_i\in \{0,1\}$. 

Putting the above equations together, Eq.~\eqref{eq: hierarchical likelihood counts} implies (where we canceled the finite size volume element $\prod_{i=1}^{n_C}\finiteData{i}$ on both sides)
\begin{equation}
\label{eq: hierarchical likelihood data finite vol}
    \p(\{\Data_i\}_{c\in C}|\hyperparams, \densityDM)
    =
    \frac{e^{-\sum_{i=1}^N\mu(\Data_i|\hyperparams,\densityDM)\,\finiteData{i}}}{n_C!}\prod_{i=1}^{n_C}
    \rateAll{i}
    \,.
\end{equation}
Taking the discretization to the infinitesimal limit as discussed in the beginning of the section, we recover \cite{Mandel:2018mve}
\begin{equation}
\label{eq: full hierarchical likelihood}
    \p(\{\Data\}_{c\in C}|\hyperparams, \densityDM)
    =
    \frac{e^{-\int  \mu(\Data|\hyperparams,\densityDM)\,\dd \Data}}{n_C!}\prod_{i=1}^{n_C}
    \rateAll{i}
    \,,
\end{equation}
the expression that describes the hierarchical likelihood of observing a set of observations $\{\Data\}_{c\in C}$ given hyperparameters and a \gls{dm} density contrast.

\section{Derivation of a maximum likelihood estimator for the true number count}
\label{app: derivation maximum likelihood estimator for the true number count}

In the following we derive an approximation for the maximum likelihood estimator for the true galaxy number count, given a DM density field, an absolute rate $\rateAbspixel$ in voxel I, and hyperparameters $\hyperparams$.

Starting from Eq.~\eqref{eq: likelihood data with ng depedency}, we absorb any variables independent of $\countGalaxiesTruepixel$ in a constant, we have
\begin{equation}
    \p(\Data_{c\in C_{\rm det}}, \countGalaxiesTruepixelall|\hyperparams,\densityDM) = \left[{\rm constant}\right] 
    \prod_{I\in\mathcal{I}}
    \Biggl\{
    {\countGalaxiesTruepixel \choose \countGalaxiesObspixel}
    \frac{\rateAbspixel^{\countGalaxiesTruepixel}}{\countGalaxiesTruepixel!}
    \left[
    \pnotdetpixel
    \right] ^ {\countGalaxiesTruepixel}
    \Biggr\}
    \,.
\end{equation}
We will compute the maximum likelihood for a galaxy number count in one specific voxel $I$, allowing us to absorb all other terms of the product $\prod_{J\in\mathcal{I},J\neq I}$ in the constant (as well as all terms independent of $\countGalaxiesTruepixel$), leading to 
\begin{equation}
    \p(\Data_{c\in C_{\rm det}}, \countGalaxiesTruepixel|\hyperparams,\densityDM) = \left[{\rm constant}\right] 
    e^{\countGalaxiesTruepixel} (\countGalaxiesTruepixel-\countGalaxiesObspixel)^{\countGalaxiesObspixel-\countGalaxiesTruepixel-\frac{1}{2}}
    \rateAbspixel^{\countGalaxiesTruepixel}
    \left[
        \pnotdetpixel
    \right] ^ {\countGalaxiesTruepixel}
    \,,
\end{equation}
where we have used Stirling's approximation $n!\approx \sqrt{2 \pi n}\left(\frac{n}{e}\right)^n $ which is valid for large $n$.
Hence, the above expression only applies for $(\countGalaxiesTruepixel-\countGalaxiesObspixel)\gg 1$. 
To obtain the maximum likelihood estimator for $\countGalaxiesTruepixel$, we solve $\partial_{\countGalaxiesTruepixel}(\p(\Data_{c\in C_{\rm det}}, \countGalaxiesTruepixel|\hyperparams,\densityDM)) = 0$, which gives
\begin{equation}
\label{eq: Stirling maximum likelihood estimator}
    \text{Maximum likelihood}(\countGalaxiesTruepixel)
    =
    \countGalaxiesObspixel+\rateAbspixel\,\pnotdetpixel e^{W\left(-\frac{1}{2 \rateAbspixel\,\pnotdetpixel}\right)}
    \,,
\end{equation}
with $W(z)$ the product log, that solves the equation $z = W e^W$.
We will now assume that the argument of the product log is small, allowing us to use the perturbative solution of 
\begin{equation}
    W(\epsilon) = \epsilon 
    - \epsilon ^ 2 
    +\frac{3}{2}\epsilon ^ 3 
    + \mathcal{O}(\epsilon^4) \,,
\end{equation}
that is only valid for $\epsilon<0$. Using this, and Taylor-expanding the exponential in Eq.~\eqref{eq: Stirling maximum likelihood estimator} gives
\begin{equation}
\label{eq: maximum likelihood estimator for true number count approximation for large rates}
     \text{Maximum likelihood}(\countGalaxiesTruepixel)
    =
    \countGalaxiesObspixel - \frac{1}{2} + \rateAbspixel\, \pnotdetpixel - 
    \frac{1}{8}\frac{1}{\rateAbspixel\,\pnotdetpixel}
    +
    \mathcal{O}\left((\rateAbspixel\,\pnotdetpixel)^{-2}\right)
    \,.
\end{equation}
As expected, the estimate for the true number of galaxies is sum of the observed number of galaxies, $\countGalaxiesObspixel$, and the overall galaxy number rate weighed by the probability of non-detection. 
The alert reader might be surprised to find no dependency of the estimator on the surrounding observed or true number counts. Note however, that this expression was derived by holding the true number counts of all other voxels constant.

\section{Prior range}
\label{app: prior range}

We summarize in Table~\ref{tab: priors} the prior ranges we consider.  

\begin{table}
\centering
\begin{tabular}{lll}
\hline
\textbf{Variable} & \textbf{Name} & \textbf{Prior} \\
\hline
\hline
\multicolumn{3}{c}{\textbf{Cosmological Parameters}} \\
\hline
$H_0$ & Hubble constant & $\mathcal{U}(64,84)$ \\
$\omegab$ & Baryon density & $\mathcal{U}(0.0185,0.026)$ \\
$\omegacdm$ & Cold dark matter density & $\mathcal{U}(0.05,0.25)$ \\
\hline
\multicolumn{3}{c}{\textbf{Power Spectrum Parameters}} \\
\hline
$\cmin$ & Halo bias parameter & $\mathcal{U}(2.1,3.9)$ \\
$\etazero$ & Power spectrum parameter & $\mathcal{U}(0.55,0.95)$ \\
$\ln10^{10}A_s$ & Power spectrum amplitude & $\mathcal{U}(1.61,3.9)$ \\
$n_s$ & Spectral index & $\mathcal{U}(0.83,1.09)$ \\
$\PSamplitude$ & Phenom amplitude & $\mathcal{U}(800,8000)^*$ \\
$\PSn$ & Phenom spectral index & $\mathcal{U}(-0.8,0.3)^*$ \\
$\PSalpha$ & Phenom running spectral index & $-0.37$ \\
\hline
\multicolumn{3}{c}{\textbf{Halo Bias Model Parameters}} \\
\hline
$\gamma$ & Galaxy bias parameter & $\mathcal{U}(0.5,2.5)$ \\
$\epsilong$ & Galaxy bias parameter & $\mathcal{U}(0.5,1.4)$ \\
$\deltagexp$ & Exponential cut-off & $\mathcal{U}(-0.8,0.2)$ \\
$\rhogbar$ & Galaxy density & $\mathcal{U}(0.01,0.2)$ \\
\hline
\multicolumn{3}{c}{\textbf{Magnitude Model Parameters}} \\
\hline
$\mu$ & Magnitude parameter & $\mathcal{U}(97.924,105.924)$ \\
$\sigma$ & Magnitude parameter & $\mathcal{U}(165.502,225.502)$ \\
$\epsilon$ & Magnitude parameter & $\mathcal{U}(-3.172,-1.772)$ \\
$\epsilon_2$ & Magnitude parameter & $\mathcal{U}(0.019,0.099)$ \\
$\muFraction$ & Faint fraction & $\mathcal{U}(0.2,0.5)$ \\
$\muThresholdAbove$ & Absolute magnitude for $\muFraction$ & -17 \\ 
$p_{\rm m,a,0}$ & Magnitude parameter & $\mathcal{U}(-0.391,-0.111)$ \\
$p_{\rm m,a,1}$ & Magnitude parameter & $\mathcal{U}(-0.128,0.152)$ \\
$p_{\rm m,a,2}$ & Magnitude parameter & $\mathcal{U}(0.172,0.452)$ \\
$p_{\rm m,a,3}$ & Magnitude parameter & $\mathcal{U}(-0.142,0.138)$ \\
$p_{\rm m,a,4}$ & Magnitude parameter & $\mathcal{U}(0.046,0.326)$ \\
$p_{\rm m,a,5}$ & Magnitude parameter & $\mathcal{U}(-0.091,0.189)$ \\
$p_{\rm m,b,0}$ & Magnitude parameter & $\mathcal{U}(0.078,0.358)$ \\
$p_{\rm m,b,1}$ & Magnitude parameter & $\mathcal{U}(0.023,0.303)$ \\
$p_{\rm m,b,2}$ & Magnitude parameter & $\mathcal{U}(0.397,0.677)$ \\
$p_{\rm m,b,3}$ & Magnitude parameter & $\mathcal{U}(10^{-6},0.160)$ \\
$p_{\rm m,b,4}$ & Magnitude parameter & $\mathcal{U}(0.269,0.549)$ \\
$p_{\rm m,b,5}$ & Magnitude parameter & $\mathcal{U}(0.259,0.539)$ \\
\hline
\end{tabular}
\caption{Summary of priors used in the analysis. We denote uniform priors with $\mathcal{U}$. For a visualization of the different magnitude model spanned by the prior ranges, consider Fig.~\ref{fig: example magnitude model}. The asterisk indicates that the prior of $\PSamplitude$ is varied: for the flat power spectrum model we adapt a more agnostic prior of $\mathcal{U}(8,8000)$. For the case that analyzes data including redshift uncertainties, we increase the prior of $\PSn$ to $\mathcal{U}(-2,0.3)$. }
\label{tab: priors}
\end{table}

\bibliographystyle{unsrt}
\bibliography{references}

\begin{thebibliography}{10}

\bibitem{LIGOScientific:2017vwq}
B.~P. Abbott et~al.
\newblock {GW170817: Observation of Gravitational Waves from a Binary Neutron Star Inspiral}.
\newblock {\em Phys. Rev. Lett.}, 119(16):161101, 2017.

\bibitem{LIGOScientific:2017adf}
B.~P. Abbott et~al.
\newblock {A gravitational-wave standard siren measurement of the Hubble constant}.
\newblock {\em Nature}, 551(7678):85--88, 2017.

\bibitem{Schutz:1986gp}
Bernard~F. Schutz.
\newblock {Determining the Hubble Constant from Gravitational Wave Observations}.
\newblock {\em Nature}, 323:310--311, 1986.

\bibitem{DelPozzo:2011vcw}
Walter Del~Pozzo.
\newblock {Inference of the cosmological parameters from gravitational waves: application to second generation interferometers}.
\newblock {\em Phys. Rev. D}, 86:043011, 2012.

\bibitem{Gray:2019ksv}
Rachel Gray et~al.
\newblock {Cosmological inference using gravitational wave standard sirens: A mock data analysis}.
\newblock {\em Phys. Rev. D}, 101(12):122001, 2020.

\bibitem{Gray:2021sew}
Rachel Gray, Chris Messenger, and John Veitch.
\newblock {A pixelated approach to galaxy catalogue incompleteness: improving the dark siren measurement of the Hubble constant}.
\newblock {\em Mon. Not. Roy. Astron. Soc.}, 512(1):1127--1140, 2022.

\bibitem{Finke:2021aom}
Andreas Finke, Stefano Foffa, Francesco Iacovelli, Michele Maggiore, and Michele Mancarella.
\newblock {Cosmology with LIGO/Virgo dark sirens: Hubble parameter and modified gravitational wave propagation}.
\newblock {\em JCAP}, 08:026, 2021.

\bibitem{Turski:2023lxq}
Cezary Turski, Maciej Bilicki, Gergely D\'alya, Rachel Gray, and Archisman Ghosh.
\newblock {Impact of modelling galaxy redshift uncertainties on the gravitational-wave dark standard siren measurement of the Hubble constant}.
\newblock {\em Mon. Not. Roy. Astron. Soc.}, 526(4):6224--6233, 2023.

\bibitem{Mastrogiovanni:2023emh}
Simone Mastrogiovanni, Danny Laghi, Rachel Gray, Giada~Caneva Santoro, Archisman Ghosh, Christos Karathanasis, Konstantin Leyde, Daniele~A. Steer, Stephane Perries, and Gregoire Pierra.
\newblock {Joint population and cosmological properties inference with gravitational waves standard sirens and galaxy surveys}.
\newblock {\em Phys. Rev. D}, 108(4):042002, 2023.

\bibitem{Gray:2023wgj}
Rachel Gray et~al.
\newblock {Joint cosmological and gravitational-wave population inference using dark sirens and galaxy catalogues}.
\newblock {\em JCAP}, 12:023, 2023.

\bibitem{Perna:2024lod}
Gabriele Perna, Simone Mastrogiovanni, and Angelo Ricciardone.
\newblock {Investigating the impact of galaxies' compact binary hosting probability for gravitational-wave cosmology}.
\newblock 5 2024.

\bibitem{Taylor:2011fs}
Stephen~R. Taylor, Jonathan~R. Gair, and Ilya Mandel.
\newblock {Hubble without the Hubble: Cosmology using advanced gravitational-wave detectors alone}.
\newblock {\em Phys. Rev. D}, 85:023535, 2012.

\bibitem{Taylor:2012db}
Stephen~R. Taylor and Jonathan~R. Gair.
\newblock {Cosmology with the lights off: standard sirens in the Einstein Telescope era}.
\newblock {\em Phys. Rev. D}, 86:023502, 2012.

\bibitem{Farr:2019twy}
Will~M. Farr, Maya Fishbach, Jiani Ye, and Daniel Holz.
\newblock {A Future Percent-Level Measurement of the Hubble Expansion at Redshift 0.8 With Advanced LIGO}.
\newblock {\em Astrophys. J. Lett.}, 883(2):L42, 2019.

\bibitem{Mastrogiovanni:2021wsd}
S.~Mastrogiovanni, K.~Leyde, C.~Karathanasis, E.~Chassande-Mottin, D.~A. Steer, J.~Gair, A.~Ghosh, R.~Gray, S.~Mukherjee, and S.~Rinaldi.
\newblock {On the importance of source population models for gravitational-wave cosmology}.
\newblock {\em Phys. Rev. D}, 104(6):062009, 2021.

\bibitem{Mancarella:2021ecn}
Michele {Mancarella}, Edwin {Genoud-Prachex}, and Michele {Maggiore}.
\newblock {Cosmology and modified gravitational wave propagation from binary black hole population models}.
\newblock {\em PhRvD}, 105(6):064030, March 2022.

\bibitem{Leyde:2022orh}
Konstantin Leyde, Simone Mastrogiovanni, Dani\`ele~A. Steer, Eric Chassande-Mottin, and Christos Karathanasis.
\newblock {Current and future constraints on cosmology and modified gravitational wave friction from binary black holes}.
\newblock {\em JCAP}, 09:012, 2022.

\bibitem{Ezquiaga:2022zkx}
Jose~Mar\'\i{}a Ezquiaga and Daniel~E. Holz.
\newblock {Spectral Sirens: Cosmology from the Full Mass Distribution of Compact Binaries}.
\newblock {\em Phys. Rev. Lett.}, 129(6):061102, 2022.

\bibitem{Pierra:2023deu}
Gr\'egoire Pierra, Simone Mastrogiovanni, St\'ephane Perri\`es, and Michela Mapelli.
\newblock {Study of systematics on the cosmological inference of the Hubble constant from gravitational wave standard sirens}.
\newblock {\em Phys. Rev. D}, 109(8):083504, 2024.

\bibitem{Leyde:2023iof}
Konstantin Leyde, Stephen~R. Green, Alexandre Toubiana, and Jonathan Gair.
\newblock {Gravitational wave populations and cosmology with neural posterior estimation}.
\newblock {\em Phys. Rev. D}, 109(6):064056, 2024.

\bibitem{Farah:2024xub}
Amanda~M. Farah, Thomas~A. Callister, Jose~Mar\'\i{}a Ezquiaga, Michael Zevin, and Daniel~E. Holz.
\newblock {No need to know: astrophysics-free gravitational-wave cosmology}.
\newblock 4 2024.

\bibitem{MaganaHernandez:2024uty}
Ignacio Maga\~na Hernandez and Anarya Ray.
\newblock {Beyond Gaps and Bumps: Spectral Siren Cosmology with Non-Parametric Population Models}.
\newblock 4 2024.

\bibitem{Dalya:2018cnd}
Gergely D\'alya, G\'abor Galg\'oczi, L\'aszl\'o Dobos, Zsolt Frei, Ik~Siong Heng, Ronaldas Macas, Christopher Messenger, P\'eter Raffai, and Rafael~S. de~Souza.
\newblock {GLADE: A galaxy catalogue for multimessenger searches in the advanced gravitational-wave detector era}.
\newblock {\em Mon. Not. Roy. Astron. Soc.}, 479(2):2374--2381, 2018.

\bibitem{Dalya:2021ewn}
G.~D\'alya et~al.
\newblock {GLADE+~: an extended galaxy catalogue for multimessenger searches with advanced gravitational-wave detectors}.
\newblock {\em Mon. Not. Roy. Astron. Soc.}, 514(1):1403--1411, 2022.

\bibitem{Dalang:2023ehp}
Charles Dalang and Tessa Baker.
\newblock {The clustering of dark sirens' invisible host galaxies}.
\newblock {\em JCAP}, 02:024, 2024.

\bibitem{det1-aligo2015}
J.~Aasi et~al.
\newblock Advanced {LIGO}.
\newblock {\em Classical and Quantum Gravity}, 32(7):074001, mar 2015.

\bibitem{det2-aLIGO:2020wna}
Aaron Buikema et~al.
\newblock {Sensitivity and performance of the Advanced LIGO detectors in the third observing run}.
\newblock {\em Phys. Rev. D}, 102(6):062003, 2020.

\bibitem{det3-Tse:2019wcy}
M.~Tse et~al.
\newblock {Quantum-Enhanced Advanced LIGO Detectors in the Era of Gravitational-Wave Astronomy}.
\newblock {\em Phys. Rev. Lett.}, 123(23):231107, 2019.

\bibitem{det4-VIRGO:2014yos}
F.~Acernese et~al.
\newblock {Advanced Virgo: a second-generation interferometric gravitational wave detector}.
\newblock {\em Class. Quant. Grav.}, 32(2):024001, 2015.

\bibitem{det5-Virgo:2019juy}
F.~Acernese et~al.
\newblock {Increasing the Astrophysical Reach of the Advanced Virgo Detector via the Application of Squeezed Vacuum States of Light}.
\newblock {\em Phys. Rev. Lett.}, 123(23):231108, 2019.

\bibitem{Sathyaprakash:2009xt}
B.~S. Sathyaprakash, B.~F. Schutz, and C.~Van Den~Broeck.
\newblock {Cosmography with the Einstein Telescope}.
\newblock {\em Class. Quant. Grav.}, 27:215006, 2010.

\bibitem{Chan:2018csa}
Man~Leong Chan, Chris Messenger, Ik~Siong Heng, and Martin Hendry.
\newblock {Binary Neutron Star Mergers and Third Generation Detectors: Localization and Early Warning}.
\newblock {\em Phys. Rev. D}, 97(12):123014, 2018.

\bibitem{Maggiore:2019uih}
Michele Maggiore et~al.
\newblock {Science Case for the Einstein Telescope}.
\newblock {\em JCAP}, 03:050, 2020.

\bibitem{Evans:2021gyd}
Matthew Evans et~al.
\newblock {A Horizon Study for Cosmic Explorer: Science, Observatories, and Community}.
\newblock 9 2021.

\bibitem{Branchesi:2023mws}
Marica Branchesi et~al.
\newblock {Science with the Einstein Telescope: a comparison of different designs}.
\newblock {\em JCAP}, 07:068, 2023.

\bibitem{Muttoni:2023prw}
Niccol\`o Muttoni, Danny Laghi, Nicola Tamanini, Sylvain Marsat, and David Izquierdo-Villalba.
\newblock {Dark siren cosmology with binary black holes in the era of third-generation gravitational wave detectors}.
\newblock 3 2023.

\bibitem{Chen:2024gdn}
Hsin-Yu Chen, Jose~Mar\'\i{}a Ezquiaga, and Ish Gupta.
\newblock {Cosmography with next-generation gravitational wave detectors}.
\newblock {\em Class. Quant. Grav.}, 41(12):125004, 2024.

\bibitem{ligo_observation_run_o4_2023}
David Shoemaker, Alessio Rocchi, and Shinji Miyoki.
\newblock Observing scenario timeline graphic, post-o3.
\newblock \url{https://dcc.ligo.org/LIGO-G2002127/public}, 2024.
\newblock {Accessed on September 22, 2024}.

\bibitem{Jasche:2009hx}
J.~Jasche, F.~S. Kitaura, B.~D. Wandelt, and T.~A. Ensslin.
\newblock {Bayesian power-spectrum inference for Large Scale Structure data}.
\newblock {\em Mon. Not. Roy. Astron. Soc.}, 406(1):60--85, 2010.

\bibitem{Jasche:2009hz}
J.~Jasche and F.~S. Kitaura.
\newblock {Fast Hamiltonian sampling for large scale structure inference}.
\newblock {\em Mon. Not. Roy. Astron. Soc.}, 407:29, 2010.

\bibitem{Jasche:2009ia}
J.~Jasche, F.~S. Kitaura, C.~Li, and T.~A. Ensslin.
\newblock {Bayesian non-linear large scale structure inference of the Sloan Digital Sky Survey data release 7}.
\newblock {\em Mon. Not. Roy. Astron. Soc.}, 409:355, 2010.

\bibitem{Jasche:2012kq}
Jens Jasche and Benjamin~D. Wandelt.
\newblock {Bayesian physical reconstruction of initial conditions from large scale structure surveys}.
\newblock {\em Mon. Not. Roy. Astron. Soc.}, 432:894, 2013.

\bibitem{Jasche:2013lwa}
Jens Jasche and Benjamin~D. Wandelt.
\newblock {Methods for Bayesian power spectrum inference with galaxy surveys}.
\newblock {\em Astrophys. J.}, 779:15, 2013.

\bibitem{Ramanah:2018eed}
Doogesh~Kodi Ramanah, Guilhem Lavaux, Jens Jasche, and Benjamin~D. Wandelt.
\newblock {Cosmological inference from Bayesian forward modelling of deep galaxy redshift surveys}.
\newblock {\em Astron. Astrophys.}, 621:A69, 2019.

\bibitem{He:2018ggn}
Siyu He, Yin Li, Yu~Feng, Shirley Ho, Siamak Ravanbakhsh, Wei Chen, and Barnab\'as P\'oczos.
\newblock {Learning to Predict the Cosmological Structure Formation}.
\newblock {\em Proc. Nat. Acad. Sci.}, 116(28):13825--13832, 2019.

\bibitem{Kaushal:2021hqv}
Neerav Kaushal, Francisco Villaescusa-Navarro, Elena Giusarma, Yin Li, Conner Hawry, and Mauricio Reyes.
\newblock {NECOLA: Toward a Universal Field-level Cosmological Emulator}.
\newblock {\em Astrophys. J.}, 930(2):115, 2022.

\bibitem{Jamieson:2022lqc}
Drew Jamieson, Yin Li, Renan~Alves de~Oliveira, Francisco Villaescusa-Navarro, Shirley Ho, and David~N. Spergel.
\newblock {Field-level Neural Network Emulator for Cosmological N-body Simulations}.
\newblock {\em Astrophys. J.}, 952(2):145, 2023.

\bibitem{SimBIG:2023ywd}
Pablo Lemos et~al.
\newblock {Field-level simulation-based inference of galaxy clustering with convolutional neural networks}.
\newblock {\em Phys. Rev. D}, 109(8):083536, 2024.

\bibitem{deSanti:2023rsw}
Natal\'\i{} S.~M. de~Santi et~al.
\newblock {Field-level simulation-based inference with galaxy catalogs: the impact of systematic effects}.
\newblock 10 2023.

\bibitem{Saadeh:2024vuj}
Daniela Saadeh, Kazuya Koyama, and Xan Morice-Atkinson.
\newblock {A field-level emulator for modified gravity}.
\newblock 6 2024.

\bibitem{Dalang:2024gfk}
Charles Dalang, Bartolomeo Fiorini, and Tessa Baker.
\newblock {Large scale structure prior knowledge in the dark siren method}.
\newblock 10 2024.

\bibitem{Agrawal:2017khv}
Aniket Agrawal, Ryu Makiya, Chi-Ting Chiang, Donghui Jeong, Shun Saito, and Eiichiro Komatsu.
\newblock {Generating Log-normal Mock Catalog of Galaxies in Redshift Space}.
\newblock {\em JCAP}, 10:003, 2017.

\bibitem{2003moco.book.....D}
Scott {Dodelson}.
\newblock {\em {Modern Cosmology}}.
\newblock 2003.

\bibitem{SpurioMancini:2021ppk}
Alessio Spurio~Mancini, Davide Piras, Justin Alsing, Benjamin Joachimi, and Michael~P. Hobson.
\newblock {CosmoPower: emulating cosmological power spectra for accelerated Bayesian inference from next-generation surveys}.
\newblock {\em Mon. Not. Roy. Astron. Soc.}, 511(2):1771--1788, 2022.

\bibitem{Piras:2023aub}
D.~Piras and A.~Spurio~Mancini.
\newblock {CosmoPower-JAX: high-dimensional Bayesian inference with differentiable cosmological emulators}.
\newblock 5 2023.

\bibitem{Blas:2011rf}
Diego Blas, Julien Lesgourgues, and Thomas Tram.
\newblock {The Cosmic Linear Anisotropy Solving System (CLASS) II: Approximation schemes}.
\newblock {\em JCAP}, 07:034, 2011.

\bibitem{Mead:2015yca}
Alexander Mead, John Peacock, Catherine Heymans, Shahab Joudaki, and Alan Heavens.
\newblock {An accurate halo model for fitting non-linear cosmological power spectra and baryonic feedback models}.
\newblock {\em Mon. Not. Roy. Astron. Soc.}, 454(2):1958--1975, 2015.

\bibitem{Mead:2016zqy}
Alexander Mead, Catherine Heymans, Lucas Lombriser, John Peacock, Olivia Steele, and Hans Winther.
\newblock {Accurate halo-model matter power spectra with dark energy, massive neutrinos and modified gravitational forces}.
\newblock {\em Mon. Not. Roy. Astron. Soc.}, 459(2):1468--1488, 2016.

\bibitem{Coles:1991if}
Peter Coles and Bernard Jones.
\newblock {A Lognormal model for the cosmological mass distribution}.
\newblock {\em Mon. Not. Roy. Astron. Soc.}, 248:1--13, 1991.

\bibitem{Bertone:2007sj}
Serena Bertone, Gabriella De~Lucia, and Peter~A. Thomas.
\newblock {The recycling of gas and metals in galaxy formation: Predictions of a dynamical feedback model}.
\newblock {\em Mon. Not. Roy. Astron. Soc.}, 379:1143--1154, 2007.

\bibitem{Springel:2005nw}
Volker Springel et~al.
\newblock {Simulating the joint evolution of quasars, galaxies and their large-scale distribution}.
\newblock {\em Nature}, 435:629--636, 2005.

\bibitem{Kuhlen:2012ft}
Michael Kuhlen, Mark Vogelsberger, and Raul Angulo.
\newblock {Numerical Simulations of the Dark Universe: State of the Art and the Next Decade}.
\newblock {\em Phys. Dark Univ.}, 1:50--93, 2012.

\bibitem{Reddick:2012qy}
Rachel~M. Reddick, Risa~H. Wechsler, Jeremy~L. Tinker, and Peter~S. Behroozi.
\newblock {The Connection between Galaxies and Dark Matter Structures in the Local Universe}.
\newblock {\em Astrophys. J.}, 771:30, 2013.

\bibitem{Zheng:2015iia}
Zheng Zheng and Hong Guo.
\newblock {Accurate and Efficient Halo-based Galaxy Clustering Modelling with Simulations}.
\newblock {\em Mon. Not. Roy. Astron. Soc.}, 458(4):4015--4024, 2016.

\bibitem{Chaves-Montero:2015iga}
Jon\'as Chaves-Montero, Raul~E. Angulo, Joop Schaye, Matthieu Schaller, Robert~A. Crain, Michelle Furlong, and Tom Theuns.
\newblock {Subhalo abundance matching and assembly bias in the EAGLE simulation}.
\newblock {\em Mon. Not. Roy. Astron. Soc.}, 460(3):3100--3118, 2016.

\bibitem{Zavala:2019gpq}
Jes\'us Zavala and Carlos~S. Frenk.
\newblock {Dark matter haloes and subhaloes}.
\newblock {\em Galaxies}, 7(4):81, 2019.

\bibitem{deMartino:2020gfi}
Ivan de~Martino, Sankha~S. Chakrabarty, Valentina Cesare, Arianna Gallo, Luisa Ostorero, and Antonaldo Diaferio.
\newblock {Dark matters on the scale of galaxies}.
\newblock {\em Universe}, 6(8):107, 2020.

\bibitem{Kokron:2021xgh}
Nickolas Kokron, Joseph DeRose, Shi-Fan Chen, Martin White, and Risa~H. Wechsler.
\newblock {The cosmology dependence of galaxy clustering and lensing from a hybrid N-body\textendash{}perturbation theory model}.
\newblock {\em Mon. Not. Roy. Astron. Soc.}, 505(1):1422--1440, 2021.

\bibitem{Neyrinck:2013ezr}
Mark~C. Neyrinck, Miguel~A. Aragon-Calvo, Donghui Jeong, and Xin Wang.
\newblock {A halo bias function measured deeply into voids without stochasticity}.
\newblock {\em Mon. Not. Roy. Astron. Soc.}, 441(1):646--655, 2014.

\bibitem{LSST:2008ijt}
\v{Z}eljko Ivezi\'c et~al.
\newblock {LSST: from Science Drivers to Reference Design and Anticipated Data Products}.
\newblock {\em Astrophys. J.}, 873(2):111, 2019.

\bibitem{Euclid:2021icp}
R.~Scaramella et~al.
\newblock {Euclid preparation. I. The Euclid Wide Survey}.
\newblock {\em Astron. Astrophys.}, 662:A112, 2022.

\bibitem{Mandel:2018mve}
Ilya Mandel, Will~M. Farr, and Jonathan~R. Gair.
\newblock {Extracting distribution parameters from multiple uncertain observations with selection biases}.
\newblock {\em Mon. Not. Roy. Astron. Soc.}, 486(1):1086--1093, 2019.

\bibitem{Neal:2011mrf}
Radford~M. Neal.
\newblock {\em {Handbook of Markov Chain Monte Carlo}}.
\newblock 5 2011.

\bibitem{Hoffman:2011ukg}
Matthew~D. Hoffman and Andrew Gelman.
\newblock {The No-U-Turn Sampler: Adaptively Setting Path Lengths in Hamiltonian Monte Carlo}.
\newblock 11 2011.

\bibitem{Betancourt:2017ebh}
Michael Betancourt.
\newblock {A Conceptual Introduction to Hamiltonian Monte Carlo}.
\newblock 1 2017.

\bibitem{Phan:2019elc}
Du~Phan, Neeraj Pradhan, and Martin Jankowiak.
\newblock {Composable Effects for Flexible and Accelerated Probabilistic Programming in NumPyro}.
\newblock 12 2019.

\bibitem{coleman_krawczyk_2024_12167630}
Coleman Krawczyk.
\newblock Ckrawczyk/multihmcgibbs: v1.0.0, June 2024.

\bibitem{Gelman:1992zz}
Andrew Gelman and Donald~B. Rubin.
\newblock {Inference from Iterative Simulation Using Multiple Sequences}.
\newblock {\em Statist. Sci.}, 7:457--472, 1992.

\bibitem{LIGOScientific:2021aug}
R.~Abbott et~al.
\newblock {Constraints on the Cosmic Expansion History from GWTC\textendash{}3}.
\newblock {\em Astrophys. J.}, 949(2):76, 2023.

\bibitem{Vijaykumar:2023bgs}
Aditya Vijaykumar, Maya Fishbach, Susmita Adhikari, and Daniel~E. Holz.
\newblock {Inferring Host-galaxy Properties of LIGO\textendash{}Virgo\textendash{}KAGRA\textquoteright{}s Black Holes}.
\newblock {\em Astrophys. J.}, 972(2):157, 2024.

\bibitem{Artale:2019tfl}
M.~Celeste Artale, Michela Mapelli, Yann Bouffanais, Nicola Giacobbo, Mario Spera, and Mario Pasquato.
\newblock {Mass and star formation rate of the host galaxies of compact binary mergers across cosmic time}.
\newblock {\em Mon. Not. Roy. Astron. Soc.}, 491(3):3419--3434, 2020.

\bibitem{EUCLID:2011zbd}
R.~Laureijs et~al.
\newblock {Euclid Definition Study Report}.
\newblock 10 2011.

\bibitem{jax2018github}
James Bradbury, Roy Frostig, Peter Hawkins, Matthew~James Johnson, Chris Leary, Dougal Maclaurin, George Necula, Adam Paszke, Jake Vander{P}las, Skye Wanderman-{M}ilne, and Qiao Zhang.
\newblock {JAX}: composable transformations of {P}ython+{N}um{P}y programs, 2018.

\end{thebibliography}

\end{document}